\documentclass[reprint,amsmath,amssymb, prb, aps, showpacs]{revtex4-1}

\usepackage{amssymb,amsmath,color,amsthm}
\usepackage{hyperref}
\usepackage{tikz}
\usetikzlibrary{arrows,shapes}
\usepackage{ textcomp }

\usepackage{epsfig}
\usepackage{verbatim} 
\usepackage{subfigure} 

\newcommand{\idop}{1\!\!1}
\newcommand{\mat}{\begin{pmatrix}}
\newcommand{\tam}{\end{pmatrix}}
\newcommand{\smat}{\left(\begin{smallmatrix}}
\newcommand{\stam}{\end{smallmatrix}\right)}
\newcommand{\Integer}{\mathbb{Z}}
\DeclareMathOperator{\tr}{\text{tr}}
\DeclareMathOperator{\ad}{{\text{ad}}}
\newcommand{\g}{\mathfrak{g}}
\newcommand{\h}{\mathfrak{h}}  
\newcommand{\cB}{\mathcal{B}}
\newcommand{\cH}{\mathcal{H}}  
\newcommand{\cV}{\mathcal{V}}
\newcommand{\cW}{\mathcal{W}}
\newcommand{\fA}{\mathfrak{A}}
\definecolor{version2}{rgb}{1,0.5,0}

\begin{document}

\title{A discriminating string order parameter for\\
  topological phases of gapped $\mathbf{SU(N)}$ spin chains}
\author{Kasper Duivenvoorden}
 \email{Kasper@thp.uni-koeln.de}
\author{Thomas Quella}%
 \email{Thomas.Quella@uni-koeln.de}
 \affiliation{Institute of Theoretical Physics, University of Cologne\\
 Z\"ulpicher Stra\ss{}e 77, D-50937 Cologne, Germany 
}%

\date{\today}

\begin{abstract}
  One-dimensional gapped spin chains with symmetry
  $PSU(N)=SU(N)/\Integer_N$ are known to possess $N$ different
  topological phases. In this paper, we introduce a non-local string
  order parameter which characterizes each of these $N$ phases
  unambiguously. Numerics confirm that our order parameter allows to
  extract a quantized topological invariant from a given
  non-degenerate gapped ground state wave function. Discontinuous
  jumps in the discrete topological order that arise when varying
  physical couplings in the Hamiltonian may be used to detect quantum
  phase transitions between different topological phases.
\end{abstract}

\pacs{03.65.Vf, 75.10.Pq, 75.10.Kt}
\maketitle

% ***********************************************************************
% ***********************************************************************
% ***********************************************************************
\section{Introduction}

  For a long time, the classification of distinct phases of matter was
  synonymous with Landau's theory of symmetry breaking. The latter
  applies to systems where the symmetry of the Hamiltonian is
  spontaneously broken in the ground state, such as ferromagnets at
  sufficiently low temperature. Different phases can be
  distinguished in terms of order parameters which correspond to
  expectation values of local observables, e.g.\ the
  magnetization. Another hallmark of the theory is the existence of
  massless Goldstone modes if continuous symmetries are broken.

  Landau's paradigm was challenged with the advent of gapped physical
  systems in which distinct phases exist even though the ground state
  (or the ground states) preserves the same symmetries as the
  Hamiltonian. Such systems are characterized by topological order, a
  term that was originally coined for fractional quantum Hall
  systems. \cite{Wen:1995AdPhy..44..405W} Since topological orders
  are usually related to discrete invariants, they enjoy protection
  against continuous deformations of the system. It might happen that
  protection against deformations is not solely due to topology but
  that it rather appears in conjunction with a symmetry which has to
  be preserved. In this case one speaks about symmetry protected
  topological orders.

  The Haldane phase of $SO(3)$ invariant antiferromagnetic spin chains
  based on $S=1$ is one of the first and best understood examples of a
  non-trivial symmetry protected topological phase. For the
  interactions it is custom to choose a specific deformation of the
  Heisenberg Hamiltonian as a representative. The resulting model is
  commonly referred to as the AKLT
  chain. \cite{Affleck:PhysRevLett.59.799,Affleck:1987cy} While the
  exact ground state for the Heisenberg Hamiltonian is not known, the
  AKLT chain provides a convenient laboratory which allows to
  establish several important properties of the Haldane phase with
  full mathematical rigor. In particular, the ground states of the
  AKLT Hamiltonian are known explicitly, both for periodic and for
  open boundary conditions. Moreover, it could be proven that the
  chain has a mass gap and that ground state correlation functions of
  local observables decay exponentially.

  The evidence for the identification of the Haldane phase as a
  non-trivial topological phase of matter can be summarized as
  follows. First of all, open boundary conditions imply the existence
  of massless edge modes. The system thus exhibits a bulk-boundary
  correspondence which is widely regarded as a typical characteristic
  of non-trivial topological phases. It was later understood that the
  topological nature is due to symmetry fractionalization which allows
  the edge  modes to carry a discrete $\Integer_2$-valued topological
  quantum number. \cite{Chen:PhysRevB.83.035107,Schuch:1010.3732v3}
  Secondly, all these features can also be observed in a
  characteristic entanglement spectrum \cite{Katsura:2008JPhA...41m5304K} which
  provides a virtual realization of edges even in the presence of
  periodic boundary conditions. Finally, and most importantly for our
  present paper, there exists a non-local string order parameter,
  \cite{DenNijs:PhysRevB.40.4709} sensitive to a kind of diluted
  anti-ferromagnetic order, which allows to distinguish the
  topologically trivial from the topologically non-trivial phase.

  Various extensions of the AKLT setup to higher rank groups
  and supersymmetric systems have been considered, see e.g.\ Ref.\ 
  \onlinecite{Affleck:1987cy,Greiter:PhysRevB.75.184441,Schuricht:PhysRevB.78.014430,Tu:PhysRevB.78.094404,Arovas:PhysRevB.79.224404}.
  Other generalizations include $q$-deformations of the symmetry group
  which can be used to describe anisotropic spin chains.
  \cite{Klumper:1992ZPhyB..87..281K,Batchelor:1994IJMPB...8.3645B}
  In all these examples the matrix product (or valence bond) state
  formalism plays a crucial role.
  \cite{Fannes:1989ns,Fannes:1990px,Fannes:1990ur,Perez-Garcia:2007:MPS:2011832.2011833}
  The latter has also proven extremely useful in connection with the
  classification of symmetry protected topological phases in
  general one-dimensional spin systems.
  \cite{Chen:PhysRevB.83.035107,Schuch:1010.3732v3,Chen:PhysRevB.84.235128}
  Indeed, by now it is well known that topological phases can be
  distinguished based on the properties of (virtual) boundary modes
  that arise when the system is considered with open boundary
  conditions or when parts of the system are traced out. Matrix
  product states are relevant in this context since their boundary and
  entanglement properties are almost trivial to access. Also, there is
  a natural way to associate a so-called parent Hamiltonian to each
  matrix product state which, in turn, is realized as the ground state
  of the former.

  The classification results just mentioned yield the number of
  potential topological phases and an explicit way of constructing a
  representative Hamiltonian for each of them. However, given an
  arbitrary physical system, i.e.\ a Hilbert space, a
  Hamiltonian and a symmetry, no universal recipe how to recover its
  topological class is known at present. Since all topological
  properties are encoded in the ground state wave functions, this is
  first of all due to the lack of knowledge of the latter. But even if
  the ground states are known exactly or approximately through a
  numerical calculation, the definition of a quantity which can be
  calculated efficiently and which can discriminate between all
  different topological phases is still an open
  problem. The degeneracy of massless edge modes might serve as a
  first indication but it still leaves
  ambiguities.\cite{Tu:PhysRevB.80.014401,Pollmann:2012arXiv1204.0704P,Duivenvoorden:2012arXiv1206.2462D}
  Even access to the full entanglement spectrum (including the energy
  and all additional quantum numbers) might not be sufficient as long
  as the contributions from the two edges cannot clearly be separated
  from each other. For this reason, the most promising route to a
  complete characterization of topological phases seems to be the
  definition of suitable non-local order parameters. Important
  progress in this direction has recently been achieved in Ref.\ 
  \onlinecite{Haegeman:1201.4174v1,Pollmann:2012arXiv1204.0704P}
  (see also Ref.\ \onlinecite{PerezGarcia:PhysRevLett.100.167202,
    Nussinow:2009AnPhy.324..977N}).
  While these approaches seem to be sufficiently general to embrace
  continuous symmetry groups as well, the concrete implementations
  have mainly been concerned with discrete symmetries so far and do
  not cover the case of $PSU(N)$.

  In the present paper we will follow an alternative route and
  use it for the characterization of anti-ferromagnetic spin chains
  with $PSU(N)$ symmetry. As has been shown in Ref.\
  \onlinecite{Duivenvoorden:2012arXiv1206.2462D},  there
  are $N$ distinct topological phases which can be realized in such
  chains. These $N$ phases correspond to the $N$ different ways,
  the center $\Integer_N$ of the group $SU(N)$ can be realized on
  possible boundary spins. Just as in the $SO(3)=SU(2)/\Integer_2$
  AKLT chain before, the situation
  can be understood as a fractionalization of the physical symmetry
  $PSU(N)=SU(N)/\Integer_N$ in a setup with open boundary
  conditions. Our main result is an explicit expression for a string
  order parameter for $SU(N)$ spin chains which can easily be
  evaluated once the ground state is known, see Eq.\
  \eqref{eq:StringOrderOp}. In contrast to earlier approaches it is
  essential that our string order parameter is a matrix valued
  quantity. Instead of extracting the information about the
  topological phase from the absolute value of the matrix entries we
  will rather infer it from relative complex phases
  between off-diagonal matrix elements. It will be proven that the
  order parameter defined in this way is quantized and that it is
  sensitive to the representation class of boundary spins with respect
  to the action of $\Integer_N$. The string order parameter thus
  allows to extract a discrete topological invariant which permits to
  discriminate all $N$ distinct phases of $PSU(N)$ spin chains. It is
  important to note that the topological invariant will only change
  when the system undergoes a discontinuity. For this reason it may be
  used as a good (numerical) measure for the identification of
  topological quantum phase transitions.

  In order to check the validity and applicability of our analytical
  results we study the phase transition between two topologically
  non-trivial phases of an $PSU(3)$ spin chain. Each of the two phases
  exhibits a subtle breaking of inversion symmetry through the
  spontaneous occurrence of boundary modes. For this reason the
  Hamiltonian cannot be written as a polynomial in the invariant
  scalar product $\vec{S}_1\cdot\vec{S}_2$ but rather requires the use
  of higher order Casimir operators. To our knowledge this is the
  first time that such Casimir operators are employed systematically
  in the formulation of spin chains. We then continue with a numerical
  investigation of the topological order using DMRG. The
  quantization of the topological order and its discontinuity at the
  phase transition, see Figure \ref{fig:res}, provide a clear
  confirmation of our analytical predictions.

  Even though spin chains based on higher rank groups like
  $SU(N)$ are unlikely to be found in real materials,
  there is a chance that the corresponding Hamiltonians can be
  engineered artificially using ultracold atoms in optical lattices.
  \cite{GarciaRipoll:PhysRevLett.93.250405,2010NatPh...6..289G} Also,
  special points in the moduli space of spin chains and spin ladders
  might exhibit an enhanced symmetry. This for instance happens for
  $SO(3)$ spin chains which are known to possess an $SU(3)$ symmetric
  point for a certain choice of the couplings.\cite{Affleck:1985wb}
  It should be noted that string order parameters have also been
  suggested for other systems, e.g.\ 1D Haldane Bose insulators.
  \cite{Torre:2006PhRvL..97z0401D} Since the latter has been observed
  in experimental measurements \cite{Endres:2011Sci...334..200E} it
  seems natural that a similar experimental verification should be
  possible for $PSU(N)$ spin chains and the string order parameter
  obtained from Eq.\ \eqref{eq:StringOrderOp}.

  The paper is organized as follows. In Section \ref{sc:Preliminaries}
  we start with a concise definition of the physical setup under
  consideration and we introduce a few of the concepts that turned out
  to be useful in the classification of topological phases: Matrix
  product states and projective representations. Afterwards we provide
  a thorough discussion of the representation theory of $su(N)$ and
  review the origin of the $N$ distinct phases of $PSU(N)$ spin
  chains. Section \ref{sc:StringOrder} contains the main result of our
  paper. We introduce a string order parameter and evaluate it in the
  thermodynamic limit. In a series of arguments we show that the
  string order parameter includes discrete topological information and
  we identify the latter with the parameter specifying the topological
  phase of the spin chain. Finally, Section \ref{sc:Applications} is
  devoted to the numerical study of a family of $PSU(3)$ symmetric
  spin chains which interpolates between two topologically non-trivial
  phases. The toy model provides a clear confirmation of our
  analytical results. Some more technical parts of the proofs and a
  brief introduction into Casimir operators of $su(3)$ have been moved
  to the Appendices.

% ***********************************************************************
% ***********************************************************************
% ***********************************************************************
\section{\label{sc:Preliminaries}Preliminaries}

  In this Section we introduce the notation and the structures that
  are used in the main part of our text. We start with a description
  of the physical setup and a brief outline of the matrix product
  state formalism. The latter is used to motivate the existence of $N$
  different phases of $PSU(N)$ spin chains. We then review some
  essential aspects concerning the representation theory of the Lie
  algebra $su(N)$.

% ***********************************************************************
% ***********************************************************************
\subsection{\label{sc:Setup}Physical setup}

  Throughout this paper we are considering spin chains which are
  characterized by the following data. The spins reside at sites $k$
  on a circular chain with periodic boundary conditions, the index
  running over the set $k=1,\ldots,L$. It will be assumed
  that the length of the chain is large but finite. The spins are
  described by operators $\vec{S}_k$ which take values in the Lie
  algebra $su(N)$ and which act on on-site Hilbert spaces $\cH_k$.
  The total Hilbert space $\cH=\bigotimes_k\cH_k$ is the product of
  all on-site Hilbert spaces. For simplicity we will assume that all
  Hilbert spaces $\cH_k$ are {\em irreducible} representations of
  $su(N)$ since otherwise the system would admit a more natural
  interpretation as a {\em spin ladder} instead of a {\em spin
    chain}. Finally, the dynamics of the system is described by a
  local Hamiltonian $H$ which commutes with
  the total spin $\vec{S}=\sum_k\vec{S}_k$. It can thus be written in
  terms of Casimir operators of $su(N)$. The simplest Hamiltonians can
  be expressed as a function of $\vec{S}_k\cdot\vec{S}_l$
  (corresponding to the quadratic Casimir) where the 
  dot denotes an $su(N)$ invariant scalar product. More complicated
  Hamiltonians, e.g.\ involving many-body interactions or breaking
  the permutation symmetry between the two sites, can be defined using
  higher order Casimir operators. An example of this type will be
  discussed in Section \ref{sc:Family}.

  Actually, the precise form of the Hamiltonian is not particularly
  important for the purpose of this paper since we will almost
  exclusively be concerned with properties of states. To be precise,
  our attention rests on the ground state $|\phi\rangle$ of the system
  which will always be assumed to be a non-degenerate finitely
  correlated state \cite{Fannes:1989ns,Fannes:1990px,Fannes:1990ur}
  (non-degenerate at least in a system with periodic boundary
  conditions). Moreover, there should exist a gap to the first excited
  state, thus implying exponential decay of local correlation
  functions. Both properties, the uniqueness and the gap, should
  persist in the thermodynamic limit.

  The simplest way to realize an anti-ferromagnetic spin chain is as
  follows. The on-site Hilbert spaces are alternating between a space
  $\cV$ and its dual $\cV^\ast$. The total Hilbert space is given by
  $\cH=(\cV\otimes\cV^\ast)^{L/2}$ and the spin dynamics is described
  by the translation invariant Heisenberg Hamiltonian
\begin{align}
  \label{eq:Heisenberg}
  H\ =\ J\sum_{i=1}^L\vec{S}_i\cdot\vec{S}_{i+1}\ \ ,
\end{align}
  with nearest neighbor interactions. The coupling constant $J$ is
  assumed to be positive, thereby favoring anti-parallel spin
  alignment. For the symmetry group $SU(2)$ and $\cV$ being the
  $S=1/2$ representation, the Hamiltonian \eqref{eq:Heisenberg} arises
  naturally from the electronic Hubbard model at half
  filling. However, with regard to the study of topological phases,
  the Heisenberg model is not ideal in many respects. First of all,
  apart from the overall normalization there are no free
  parameters in the Hamiltonian so it can only serve as a
  representative of one physical phase. Also, besides the fact that
  the ground state is not known
  exactly, the absence or presence of a gap has not been fully
  established. The absence of a gap is known for certain
  representations $\cV$. \cite{Affleck:1986pq} For other
  representations, the existence of a gap can be proven in the limit
  of ``large spin'' using a mapping to a $\sigma$-model with a
  topological $\Theta$-term.
  \cite{Haldane:PhysRevLett.50.1153,Haldane:1983464,Bykov:2012arXiv1206.2777B}
  More recently, the question of the Haldane gap has been revisited in Ref.\ 
  \onlinecite{Greiter:PhysRevB.75.184441,Rachel2010:PhysRevB.80.180420,Duivenvoorden:2012arXiv1206.2462D}.

  In order to realize different topological phases while retaining
  full analytic control over the ground state of the system, it is
  useful to consider modifications of the Heisenberg
  Hamiltonian which are obtained by generalizing the AKLT construction.
  \cite{Affleck:PhysRevLett.59.799,Affleck:1987cy} These Hamiltonians
  arise as ``parent Hamiltonians'' of specific matrix product states
  (MPS).
  \cite{Fannes:1989ns,Fannes:1990px,Fannes:1990ur,Perez-Garcia:2007:MPS:2011832.2011833}
  Since all our considerations take place on the level of
  ground states we will refrain from giving detailed expressions for
  the Hamiltonians. The only exception is a specific family of
  Hamiltonians with $su(3)$ symmetry which will be the subject of
  Section \ref{sc:Applications} and which interpolates between two
  Hamiltonians associated with different topological phases. It will
  be used to abandon the idealized environment of MPS parent
  Hamiltonians and to provide a numerical check of our ideas in a more
  realistic scenario.

% ***********************************************************************
% ***********************************************************************
\subsection{\label{sc:MPSClass}Matrix product states and topological
  phases}

  Let the vectors $|i_k\rangle$ denote an orthonormal basis of the
  on-site Hilbert spaces $\cH_k$. Using an iterated Schmidt
  decomposition, {\em any} state $|\phi\rangle$ of a periodic spin
  chain of length $L$ can be written as
  \cite{Perez-Garcia:2007:MPS:2011832.2011833}
\begin{align}
  \label{eq:MPS}
  |\phi\rangle
  \ =\ \sum_{i_1,\ldots,i_L}\tr\bigl(A^{[1]\,i_1}\cdots A^{[L]\,i_L}\bigr)
       \,|i_1\cdots i_L\rangle\ \ ,
\end{align}
  with a certain set of matrices $A^{[k]}$ carrying three different
  indices, one physical and two auxiliary ones. Such a state is known
  as a matrix product state. To be precise, one has to distinguish
  different types of MPS depending on the behavior of the system in
  the thermodynamic limit $L\to\infty$. If one wishes to describe the
  ground state of a critical system, the size of the matrices
  $A^{[k]}$ will grow beyond any limit. In our current paper we are
  only interested in gapped systems and hence we will
  assume that the dimension of the matrices $A^{[k]}$ (and their
  nature) stabilizes for sufficiently large values of $L$. The
  resulting infinite volume states are known as finitely correlated
  states. \cite{Fannes:1989ns,Fannes:1990px,Fannes:1990ur} Even though
  we are eventually interested in the thermodynamic limit, an accurate
  description of the physics of the system can be obtained by working
  with finite but large $L$ for this class of states. In the presence
  of a finite gap, there are exponential corrections to expectation
  values which quickly die away if $L$ is sufficiently large.

  The structure \eqref{eq:MPS} arises naturally if one
  associates two auxiliary spaces $\cH_{(k,L)}$ and $\cH_{(k,R)}$ to
  each physical site $k$ such that $\cH^*_{(k,R)}=\cH_{(k+1,L)}$. This
  guarantees the existence of a maximally entangled state $|I_k\rangle
  =\sum_q|q\rangle\langle q|\in\cH_{(k,R)}\otimes \cH_{(k+1,L)}$ where
  $|q\rangle$ refers to an orthonormal basis of $\cH_{(k,R)}$. The
  matrices $A^{[k]}$ can be regarded as linear maps from
  $\cH_{(k,L)}\otimes\cH_{(k,R)}$ to $\cH_k$. The state $|\phi\rangle$
  is the image of the tensor product
  $|I\rangle=|I_1\rangle\otimes\cdots\otimes|I_{L-1}\rangle$ of
  completely entangled pairs under the map
  $\fA=A^{[1]}\otimes\cdots\otimes A^{[L]}$. The application of the
  map $\fA$ to the product of completely entangled pairs $|I\rangle$
  effectively converts the tensor product into a matrix product.

  In the spin chains we are interested in, the physical Hilbert spaces
  $\cH_k$ carry a unitary representation of $SU(N)$. Moreover, the
  ground state $|\phi\rangle$ should be invariant under the action of
  $SU(N)$.\footnote{Note that invariance is implied automatically if
    the ground state is unique.} These two properties imply the
  existence of additional structures which are realized on the data of
  an MPS. Let $R^{[k]}(g)$ denote the representation of
  $SU(N)$ on the space $\cH_k$. According to
  Ref.\ \onlinecite{Sanz:PhysRevA.79.042308}, this on-site symmetry
  lifts to the auxiliary level as
\begin{align}
  \label{eq:Intertwiner}
  R^{[k]}(g)\cdot A^{[k]}
  \ =\ D^{[k]}(g)A^{[k]}D^{[k+1]}(g)^{-1}\ \ ,
\end{align}
  thereby promoting $\cH_{(k,L)}$ and $\cH_{(k,R)}$ to representations
  of $SU(N)$.\footnote{Here we implicitly assumed that
    $\cH_{(k,R)}$ and $\cH_{(k+1,L)}$ are dual not only on the level
    of vector spaces but also on the level of representations. This
    automatically implies the $SU(N)$ invariance of the completely
    entangled pairs $|I_k$\textrangle .}
  In other words, the homomorphisms $A^{[k]}$ should be equivariant
  projections from $\cH_{(k,L)}\otimes\cH_{(k,R)}$ to $\cH_k$, i.e.\ they
  should commute with the action of $SU(N)$.

  In fact, a careful inspection of relation \eqref{eq:Intertwiner}
  shows that the physical Hilbert space $\cH_k$ and the associated
  auxiliary spaces $\cH_{(k,L/R)}$ enter the discussion on a different
  footing. To understand this statement, let us for a moment assume
  that $D^{[k]}=D^{[k+1]}$ and that the auxiliary spaces form an
  {\em irreducible} representation of $SU(N)$. In view of Schur's Lemma, the
  right hand side of Eq.\ \eqref{eq:Intertwiner} -- and hence also the
  left hand side -- is invariant in this case if $g$ is chosen to be
  in the center $\Integer_N$ of the symmetry group $SU(N)$. In other
  words, $R^{[k]}$ descents to a linear representation of the quotient
  group $PSU(N)=SU(N)/\Integer_N$ while no such requirement exists for
  the matrix $D^{[k]}$. The latter only needs to implement a {\em
    projective representation} of $PSU(N)$,
\begin{align}
  D(g_1)D(g_2)\ &=\ \omega(g_1,g_2)\,D(g_1g_2)\\[2mm]\nonumber
  \text{ with }\quad
  g_1,g_2\in PSU(N) &
  \quad\text{ and }\quad
  \omega(g_1,g_2)\in U(1)\ \ ,
\end{align}
  i.e.\ a representation up to phase factors. It is known that the
  projective representations of $PSU(N)$ fall into $N$ different
  classes when considered modulo obvious equivalences (see e.g.\ Ref.\ 
  \onlinecite{Duivenvoorden:2012arXiv1206.2462D}).

  Analogous considerations apply if the assumption $D^{[k]}=D^{[k+1]}$
  fails. By choosing suitable representations of $SU(N)$ on the
  auxiliary spaces, one can realize any symmetry group $SU(N)/\Gamma$
  on the physical Hilbert spaces $\cH_k$, where
  $\Gamma\subset\Integer_N$ is an arbitrary subgroup of the center of
  $SU(N)$. It can be shown that the group $SU(N)/\Gamma$ has
  $|\Gamma|$ distinct classes of projective representations.
  \cite{Duivenvoorden:2012arXiv1206.2462D}

  In a series of papers,
  \cite{Chen:PhysRevB.83.035107,Schuch:1010.3732v3,Chen:PhysRevB.84.235128}
  the projective class of the representation of the physical symmetry
  on the auxiliary spaces
  has been identified as a topological invariant of 1D gapped spin
  chains. In other words, the projective class arising in the MPS
  representation of the respective ground states remains invariant
  upon deformation of the Hamiltonian. For the symmetry group
  $PSU(N)$, the previous argument predicts exactly $N$ distinct
  topological phases. For a general treatise on 1D spin systems with
  continuous on-site symmetries we refer the interested reader to
  Ref.\ \onlinecite{Duivenvoorden:2012arXiv1206.2462D}.

  The different topological phases of a spin chain with a given
  symmetry can all be realized explicitly by defining suitable parent
  Hamiltonians. More precisely, for each MPS $|\phi\rangle$ of the
  form \eqref{eq:MPS} there
  exists a local Hamiltonian with the following two properties:
  \cite{Perez-Garcia:2007:MPS:2011832.2011833} The
  state $|\phi\rangle$ is the unique ground state of the Hamiltonian
  and there exists a gap. When considered with open boundary
  conditions, this construction will lead to gapless edge modes which
  transform according to the projective representations
  $\cB_L=\cH_{(1,L)}$ and $\cB_R=\cH_{(L,R)}$. Even though the energy
  of boundary states will receive corrections and the degeneracy with
  the ground state might get lost upon deformation of the Hamiltonian,
  they will remain stable until the mass gap closes in the
  bulk. Intuitively, the correlation length will diverge at the phase
  transition, thus allowing the two boundaries modes of the spin chain
  to interact with each other and to disappear.

% ***********************************************************************
% ***********************************************************************
\subsection[The Lie algebra $su(N)$ and its representations]{\label{sc:SUConventions}The Lie algebra $\mathbf{su(N)}$ and its representations}

  For a more detailed discussion of $SU(N)$ spin chains and a concise
  formulation of our result we need to review the representation
  theory of the Lie algebra $su(N)$ (see e.g.\ Ref.\ 
  \onlinecite{FultonHarris:MR1153249,FuchsSchweigert}). The latter is the
  Lie algebra $\g$ of traceless $N\times N$ matrices and it is
  generated (as a vector space) by the matrices $E^{ab}$ with $a\neq
  b$ and by $H^a=E^{aa}-E^{a+1,a+1}$. Here $E^{ab}$ denotes the
  elementary matrix $(E^{ab})_{cd}=\delta_{ac}\delta_{bd}$ with a
  single non-zero entry in row $a$ and column $b$. The diagonal
  matrices $H^a$ generate the Cartan subalgebra $\h$ of $su(N)$. The
  other generators $E^{ab}$ are called positive or negative roots,
  depending on whether $a<b$ or $a>b$. As a consequence, the Lie
  algebra $su(N)$ admits a triangular decomposition
  $\g=\g_+\oplus\h\oplus\g_-$ into positive roots $\g_+$, negative
  roots $\g_-$ and the Cartan subalgebra $\h$. As a Lie algebra,
  $su(n)$ is generated by the positive and negative simple roots
  $E^{ab}$ with $|a-b|=1$.

  All finite dimensional representations $V$ of $su(N)$ are so-called
  weight representations in which all generators $H^a$ are represented
  by diagonal matrices $\rho_V(H^a)$. By abuse of notation we will
  simply omit to write the map $\rho_V$ in case it is clear that we
  are acting on a representation. If $\mu\in\h^\ast$ one says
  that a vector $v\in V$ has weight $\mu$ provided that
\begin{equation}
  H^a v  = \mu(H^a) v =: \mu_av\ \ .
\end{equation}
  The different eigenvalues $\mu_a$ can be assembled into a
  tuple $\mu=(\mu_1,\ldots,\mu_{N-1})$ of Dynkin labels and should be
  regarded as physical charges characterizing the state $v$. A
  convenient basis for the space $\h^\ast$ is given by the fundamental
  weights $\omega_a$ which are dual to the Cartan generators in the
  sense that $\omega_a(H^b) = \delta_a^b$. A weight can thus also be
  written as $\mu=\sum_a\mu_a\omega_a$. Any representation space $V$
  can be split into distinct eigenspaces with regard to the action of
  the generators $H^a$. This leads to the weight space decomposition
\begin{align}
  V\ =\ \bigoplus_{\mu\in\h^\ast}V_{\mu}\ \ .
\end{align}
  In a finite dimensional representation all weights $\mu$ are
  necessarily integral, i.e.\ $\mu_a\in\Integer$. The set of all
  weights forms the weight lattice $P$ which is an abelian group under
  addition.

  Let us now turn our attention to finite dimensional irreducible
  representations. As is well known, the latter are labeled by weights
  $\lambda$ whose Dynkin labels $\lambda_a$ are all non-negative
  integers. Such weights are called dominant. The set of dominant
  weights, denoted by $P^+$, defines the fundamental Weyl chamber of
  the weight lattice $P$. Within an irreducible representation
  $\lambda$, the different weights are all related by the application
  of roots $\alpha$. The latter should be thought of as the charges of
  the root generators $E^{ab}$ (for $a\neq b$) with respect to the
  Cartan generators $H^a$. Phrased differently, for each weight $\mu$
  in the representation $\lambda$ one has $\lambda-\mu\in Q$ where $Q$
  is the root lattice which is generated by the (finite) set of roots
  $\alpha$.

  A distinguished role is played by the adjoint representation in
  which the Lie algebra is represented on itself (regarded as a
  vector space) by means of the adjoint map
  $X\mapsto\ad_X=[X,\,\cdot\,]$. The non-zero weights of the adjoint
  representation are precisely the roots $\alpha$. The $N-1$ simple
  roots have weights $\alpha_a$ which are just the rows of the $su(N)$
  Cartan matrix $A_{ab}=2\delta_{ab}-\delta_{|a-b|,1}$. For our
  purposes it will be important that there exists a unique weight
  $\rho=\frac{1}{2}\sum_{\alpha>0}\alpha=(1,\ldots,1)$, the so-called
  Weyl vector, which has a scalar product $(\rho,\alpha_a)=1$ with
  each of the simple roots $\alpha_a$. The dual generator
  $H^\rho\in\h$ is characterized by the property
\begin{equation}
  \label{eq:aux3}
  \alpha_a(H^\rho)
  \ =\ 1
\end{equation}
  for all $a=1,\ldots,N-1$. This generator will play an important role
  in the definition of the string order parameter in Section
  \ref{sc:StringOrder}. With the previous choice of simple roots one
  can find the following explicit expression for the diagonal entries
  of the matrix $H^\rho=\text{diag}(H^\rho_1,\ldots,H_{N-1}^\rho)\in\h$, 
\begin{equation}
  H^\rho_a
  \ =\ \frac{N+1}{2}-a\ \ .
\end{equation}
  Indeed, one can easily check that this defines the unique traceless
  diagonal matrix with $H^\rho_a-H^\rho_{a+1}=1$, as required by Eq.\
  \eqref{eq:aux3}.

  The final ingredient that will be needed below is the Weyl group
  of $su(N)$. The Weyl group can be regarded as the symmetry of the
  root system. It consists of rotations and reflections which leave
  the set of roots invariant and is thus a subgroup of the orthogonal
  group in $N-1$ dimensions. For $su(N)$, the Weyl group is 
  isomorphic to the symmetric group $S_N$. Under the action of the
  Weyl group, the weight lattice $P$ may be decomposed into
  orbits. In our considerations below it will be crucial that each of
  these orbits has at least one element in the fundamental Weyl
  chamber of dominant weights $P^+$. In other words, for each weight
  $\mu\in P$ one can find an element $S\in\cW$ such that $\mu'=S\mu$
  is in the fundamental Weyl chamber, i.e.\ $\mu'\in P^+$. Note that
  the element $S$ need not be unique.

  In order to derive the explicit action of the Weyl group on a weight
  it is convenient to switch to an alternative set of elements
  $\epsilon_i$ ($i=1,\ldots,N$) which span the dual $\h^*$ of the
  Cartan algebra of $su(N)$. Given any $H\in\h$ they are defined by
  $\epsilon_i(H)=H_{ii}$. Since $su(N)$ matrices are traceless, these
  vectors satisfy the constraint $\sum_i\epsilon_i=0$ which leads to a
  slight redundancy when weights are expressed in terms of the
  $\epsilon_i$. However, this disadvantage is compensated by the
  simple transformation behavior under the action of the Weyl group
  $\cW$ which, for $su(N)$, is isomorphic to the symmetric group
  $S_N$. Given any permutation $\sigma\in S_N$ and the associated Weyl
  group element $S_\sigma\in\cW$ one simply has
\begin{equation}
  \label{eq:WeylGroupAction}
  S_\sigma(\epsilon_i)
  \ =\ \epsilon_{\sigma(i)}\ \ .
\end{equation}
  Given this formula, we can deduce the Weyl group action on any
  weight $\mu=\sum_ic_i\epsilon_i$. Note that the labels $c_i$ are, a
  priori, only defined up to a simultaneous shift. We can nevertheless
  arrive at a unique description by imposing the ``gauge''
  $\sum_ic_i=0$, and the latter will be assumed from now on. With
  these conventions the new labels are related to the standard Dynkin
  labels $\mu_a$ as
\begin{equation}
  \label{eq:WeightConversion}
  c_i
  \ =\ -\sum_{a=1}^{i-1}\frac{a}{N}\mu_a+\sum_{a=i}^{N-1}\frac{N-a}{N}\mu_a\
  \ .
\end{equation}
  This relation can be derived using the explicit form of the roots in
  terms of Dynkin labels, compare the expression for the Cartan matrix
  above.

  A second reason for using the epsilon basis is that
  $\epsilon_i(H^a)$ and $\epsilon_i(H^\rho)$ can be easily
  calculated. The former evaluates to
  $\epsilon_i(H^a)=\delta_{i,a}-\delta_{i,a+1}$, while the latter
  is $\epsilon_i(H^\rho) = \frac{N+1}{2}-i$. Note that a shift in the
  index $i$ translates directly to a shift in
  $\epsilon_i(H^\rho)$. More precisely, let $\sigma_m\in S_N$ be the
  cyclic permutation defined by $\sigma_m(i)=i+m$ (modulo $N$). Then
\begin{equation}
  \label{eq:aux4}
  \epsilon_{\sigma_m(i)}(H^\rho)
  \ =\ \epsilon_i(H^\rho) - m +N\,\theta(i+m-N)\ \ .
\end{equation}
  Here, $\theta$ is the Heaviside step function with $\theta(0)=0$.

% ***********************************************************************
% ***********************************************************************
\subsection{\label{sc:SUCase}Classes of representations}

\begin{figure}
\includegraphics[]{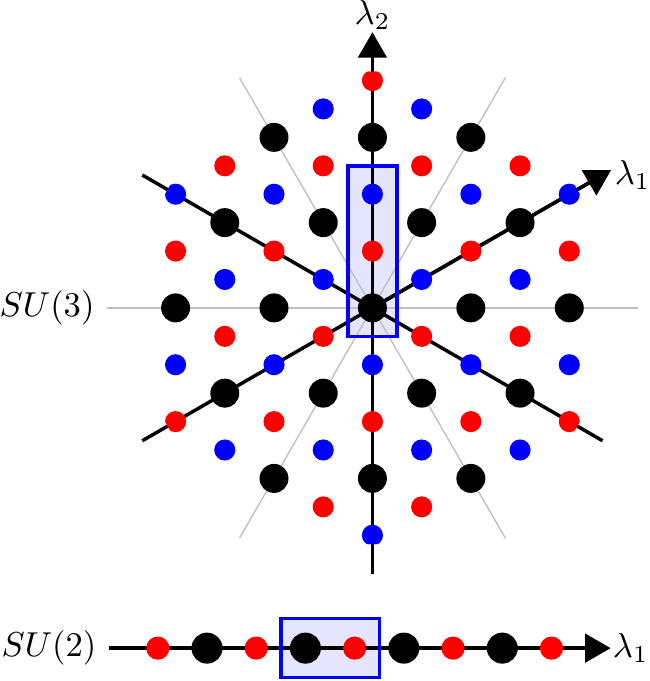}
  \caption{\label{fig:Congruence}(Color online) Visualization of
    different congruence classes for $SU(2)$ and $SU(3)$ in terms of
    colors. The shaded blue boxes are possible representatives of
    topological classes.}
\end{figure}

  As was discussed in detail in our previous article,
  \cite{Duivenvoorden:2012arXiv1206.2462D} the representations
  $\lambda$ of $su(N)$ (and hence of $SU(N)$) fall into $N$ different
  classes which can be interpreted as elements of the group $P/Q$, the
  quotient of the weight lattice $P$ by the root lattice $Q$. In terms
  of its Dynkin labels the class of the representation
  $\lambda=(\lambda_1,\ldots,\lambda_{N-1})$ is defined by
\begin{align}
  \label{eq:SUClass}
  [\lambda]
  \ \equiv\ \sum_{a=1}^{N-1}a\lambda_a\mod N\ \ .
\end{align}
  When representations are specified using Young tableaux, the class
  of a representation can be expressed as the number of boxes modulo
  $N$. \cite{Duivenvoorden:2012arXiv1206.2462D} Even though Eq.\
  \eqref{eq:SUClass} was introduced for highest weights, it can be
  extended to any weight since the expression on the right hand side
  is invariant under the action of the root lattice $Q$. In other
  words, $[\mu]=[\lambda]$ for any weight $\mu$  in a representation
  with highest weight $\lambda$. As is illustrated in Figure
  \ref{fig:Congruence}, equation \eqref{eq:SUClass} divides the weight
  lattice of $su(N)$ into $N$ different sublattices.

  The congruence class $[\lambda]$ of a representation $\lambda$
  determines whether the representation descends to quotients of
  the group $SU(N)$. More precisely, the value of $[\lambda]$
  fixes the action of the center $\Integer_N$ of $SU(N)$ on the
  representation $\lambda$. Elements of $\Integer_N\subset SU(N)$ are
  multiples $\Omega^kI_N$ of the identity matrix with $\Omega =
  \exp(\frac{2\pi i}{N})$ denoting the fundamental $N^{\text{th}}$ root
  of unity. In the representation $\lambda$, this element is mapped to
  the complex phase
  $\rho(\Omega^kI_N)=\Omega^{k[\lambda]}$. We conclude that
  representations $\lambda$ with $[\lambda]\equiv0$ are linear
  representations of $PSU(N)=SU(N)/\Integer_N$. Likewise we may ask
  whether a representation $\lambda$ lifts to any of the groups
  $SU(N)/\Integer_q$ where $\Integer_q\subset\Integer_N$ is a subgroup
  of the center. This is the case if and only if $[\lambda]\equiv0\mod
  q$ (instead of using mod $N$).
  \cite{Duivenvoorden:2012arXiv1206.2462D}

  The connection to the classification of topological phases comes in
  since representations $\lambda$ of $SU(N)$ with
  $[\lambda]\not\equiv0$ only define {\em projective} representations
  of $PSU(N)$. If the physical Hilbert spaces $\cH_k$ transform in a
  linear representation of $PSU(N)$, the (virtual) boundary spins
  might still transform in a projective representation of $PSU(N)$ as
  was discussed in Section \ref{sc:MPSClass}. The division of $SU(N)$
  representations into $N$ distinct classes which is described by Eq.\
  \eqref{eq:SUClass} in this way reflects the division of spin chains
  into $N$ distinct topological classes.

  Let us finally establish the connection to the physical spin chains
  which have been discussed in Section \ref{sc:MPSClass}. Since we
  shall be dealing with $PSU(N)$ spin chains in this paper, the
  physical Hilbert spaces $\cH_k$ (which are described by a highest
  weight $\lambda$) should all reside in the trivial class, i.e.\
  $[\cH_k]=[\lambda]\equiv0$. On the other hand, the auxiliary spaces
  $\cH_{(k,L/R)}$ can reside in non-trivial classes as long as their
  total class sums up to zero,
  $[\cH_{(k,R)}]=-[\cH_{(k,L)}]$. Together with the condition
  $[\cH_{(k,R)}]=-[\cH_{(k+1,L)}]$ which arises from the duality
  constraint $\cH_{(k,R)}^\ast=\cH_{(k+1,L)}$ this means that the
  projective class of the left and right auxiliary spaces,
  respectively, is constant all along the chain.

% ***********************************************************************
% ***********************************************************************
% ***********************************************************************
\section{\label{sc:StringOrder}A string order parameter for
  $\mathbf{SU(N)}$ spin chains}

  In this Section we introduce a non-local string order parameter for
  $SU(N)$ spin chains which reduces to the diluted anti-ferromagnetic
  order of Rommelse and den\,Nijs \cite{DenNijs:PhysRevB.40.4709} for
  $N=2$. Using transfer matrix methods we evaluate the string order
  parameter on matrix product ground states and show that it may be
  used to extract a {\em quantized} topological order parameter. The
  latter is capable of distinguishing between the $N$ different phases
  of $PSU(N)$ invariant spin chains.

% ***********************************************************************
% ***********************************************************************
\subsection{\label{sc:Definition}Definition and interpretation}

  Let $|\phi\rangle$ be the unique ground state of our spin system. We
  will assume that the system has a symmetry group $PSU(N)$ and that
  $|\phi\rangle$ is in a definite topological phase described by a
  constant $t\in\Integer_N$ (regarded as an additive group). Following
  the reasoning of Section \ref{sc:MPSClass}, the constant $t$ will be
  identified with the projective class $[\cH_{(k,R)}]$ of the right
  auxiliary representations arising in the matrix product state
  representation of $|\phi\rangle$.

  In what follows we shall prove that the ground state expectation
  value $\langle\sigma_{ij}^{ab}\rangle$ of the non-local string order
  operator
\begin{align}
  \label{eq:StringOrderOp}
  \sigma_{ij}^{ab}
  \ =\ H_i^a\exp\biggl[\frac{2\pi
    i}{N}\sum_{k=i+1}^{j-1}H_k^\rho\biggr]H_j^b
  \qquad(\text{for }i<j)
\end{align}
  contains all information required to reconstruct the value of
  $t$. It serves as a convenient tool for the measurement of the
  topological phase of the system, even in cases where the matrix
  product state representation of $|\phi\rangle$ is not known or where
  the nature of the auxiliary spaces -- regarded as a representation
  of $SU(N)$ -- is unclear. In the previous formula, $H^\rho$ refers
  to the Cartan operator associated with the Weyl vector $\rho$ (see
  Section \ref{sc:SUConventions}). For $SU(2)$, expression
  \eqref{eq:StringOrderOp} reduces to the string order
  $S_i^z\exp{(i\pi\sum S^z)}S_j^z$ introduced by Rommelse and
  den\,Nijs. \cite{DenNijs:PhysRevB.40.4709}

  In the following Section it will be proven that, in the limit
  $|i-j|\to\infty$, the dependence of the string order parameter
  $\langle\sigma_{ij}^{ab}\rangle$ on $a$ and $b$ converges
  exponentially to
\begin{equation}
\begin{split}
  \label{eq:SOLimit}
  T^{ab}
  &\ =\ \lim_{|i-j|\to\infty}\langle\sigma_{ij}^{ab}\rangle
  \ =\ C_{ij}\,\Omega^{t(a-b)}\\[2mm]
  &\qquad\qquad\text{ with }\quad
  \Omega=\exp\frac{2\pi i}{N}\ .
\end{split}
\end{equation}
  The prefactor $C_{ij}$ can be used as a first rough indication of
  whether the system resides in a topologically trivial phase or
  not. In a trivial phase we will always obtain $C_{ij}=0$ while in a
  non-trivial phase the prefactor is expected to be
  non-zero.\footnote{While we still lack a mathematical proof,
    there is numerical evidence for our assertion.} Up to this
  point, the discussion completely parallels the analysis of the
  conventional $SU(2)$
  string order. For $SU(N)$, however, the most important information
  resides in the off-diagonal entries, the complex phases
  $\Omega^{t(a-b)}$. Obviously, the constant $t$ entering this
  expression is only defined modulo $N$. In fact, as we shall see
  below, it takes values in $\Integer_N$, just as desired. It
  characterizes the projective class according to which (virtual) edge
  modes transform and it thereby determines the topological phase of the
  state $|\phi\rangle$. Whenever $C_{ij}\neq0$, the value of $t$ can
  be extracted unambiguously by calculating (or measuring) two
  different matrix elements and taking their quotient. For instance
  one immediately finds $T^{21}/T^{11}=\Omega^t$. Let us emphasize
  that a transition from one topological phase to another enforces the
  prefactor $C_{ij}$ to vanish since otherwise the parameter $t$
  cannot change its value.

  In the way it was introduced, the constant $t\in\Integer_N$
  determines the projective class of (virtual) edge modes with respect
  to the minimal quotient $PSU(N)=SU(N)/\Integer_N$ of $SU(N)$. In a
  concrete physical realization it might happen that the actual
  symmetry group is not $PSU(N)$ but rather a different quotient
  $SU(N)/\Integer_q$ where $\Integer_q\subset\Integer_N$. In this
  case, the projective classes are described by $\Integer_q$, not by
  $\Integer_N$, and $t$ has to be considered modulo $q$, see Ref.\ 
  \onlinecite{Duivenvoorden:2012arXiv1206.2462D}.

  The attentive reader may wonder why the expectation value
  \eqref{eq:SOLimit} still depends on $i$ and $j$ even after taking
  the limit $|i-j|\to\infty$. The answer is simple: The result of
  the calculation depends on the representation spaces used at sites
  $i$ and $j$ and hence on how the limit is performed. The dependence
  will disappear if the system is translation invariant.

% ***********************************************************************
% ***********************************************************************
\subsection{Evaluation}

  The proof of Eq.\ \eqref{eq:SOLimit} will proceed in two steps. We
  first prove the factorization of the matrix
  $\langle\sigma_{ij}^{ab}\rangle = \langle
  J_{i,L}^a\rangle\langle{J}_{j,R}^b\rangle$ in the thermodynamic
  limit, up to exponentially small corrections. This step uses
  transfer matrix techniques and it is
  intimately related to the matrix product state structure of
  $|\phi\rangle$. In a second step we use the Weyl symmetry of the
  weight lattice to reduce $J_{i,L}^a$ to a simpler expression. The
  latter is further analyzed in a third step from which we conclude
  that $\langle J_{i,L}^a\rangle$ depends on $a$ as $\langle
  J_{i,L}^a\rangle \propto \Omega^{at}$. The case $J_{j,R}$ can be
  dealt with analogously.

% ***********************************************************************
\subsubsection*{Step 1: Factorization}

  To prove the factorization of the matrix $T^{ab}$ we express the ground
  state $|\phi\rangle=\fA|I\rangle$ in terms of the maximally
  entangled state $|I\rangle$, see Section \ref{sc:MPSClass}.
  The possibility to write $|\phi\rangle$ in this form is a direct
  consequence of the fact that $|\phi\rangle$ can be written as a
  matrix product state. In the next step we use the intertwining
  property
\begin{align}
  H_k^a\,\fA\ =\ \fA(H_{k,L}^a+H_{k,R}^a)
\end{align}
  which expresses the physical spin operator $H_k^a$ as a sum of spin
  operators $H_{k,L}^a$ and $H_{k,R}^a$ on the two corresponding
  auxiliary sites. Using the singlet property of $|I\rangle$,
\begin{align}
  H_{k,R}^a|I\rangle\ =\ -H_{k+1,L}^a|I\rangle\ \ ,
\end{align}
  one easily sees that the phase factors in the string order operator
  $\sigma_{ij}^{ab}$ cancel out pairwise except for the two
  boundaries. We then immediately find
\begin{align}
  \langle\sigma_{ij}^{ab}\rangle
  \ =\ \frac{\langle\phi|\sigma_{ij}^{ab}\fA|I\rangle}{\langle\phi|\phi\rangle}
  \ =\ \frac{\langle\phi|\fA J_{i,L}^aJ_{j,R}^b|I\rangle}{\langle\phi|\phi\rangle}\ \ ,
\end{align}
  where the two operators $J_{i,L}^a$ and $J_{j,R}^b$ are defined by
\begin{align}\label{eq15}
  J_{i,L}^a\ &=\ (H_{i,L}^a+H_{i,R}^a)\,\Omega^{-H_{i,R}^\rho}
  \qquad\text{ and }\\[2mm]
  J_{j,R}^b\ &=\ \Omega^{-H_{j,L}^\rho}\,(H_{j,L}^b+H_{j,R}^b)\ \ .
\end{align}
  Note that each of these operators acts locally on two {\em
    auxiliary} sites. However, neither of them can be lifted to an
  operator acting locally on {\em physical} sites, i.e.\ there is no
  way to commute them back through $\fA$ without rebuilding the
  original non-local string.

\begin{figure}
\includegraphics[]{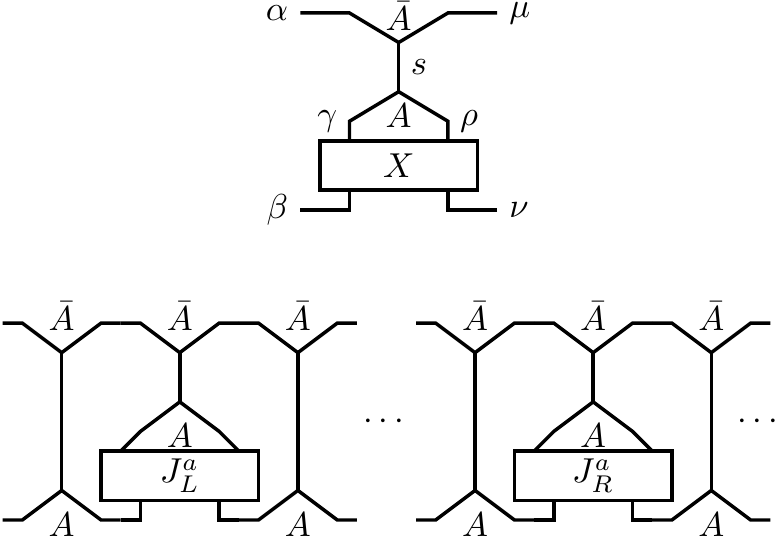}
  \caption{\label{fig:TransferMatrix}Sketch of the one-site transfer matrix
    $E_X^{[k]}$ (upper diagram) and of the expectation value
    \eqref{eq:ExptValues} (lower diagram). For the latter, periodic boundary
    conditions are assumed.}
\end{figure}

  Now that we could eliminate the non-local string connecting the two
  sites $i$ and $j$ we can evaluate the string order parameter using
  standard transfer matrix techniques. \cite{Fannes:1989ns} For that
  purpose, we write
\begin{align}
  \label{eq:ExptValues}
  \langle\phi|\fA J_{i,L}^aJ_{j,R}^b|I\rangle
  &\ =\ \\[2mm]\nonumber
 \tr\bigl(\cdots E^{[i-1]}&E^{[i]}_{J_L^a}E^{[i+1]}\cdots
  E^{[j-1]}E^{[j]}_{J_R^b}E^{[j+1]}\cdots \bigr),\\[2mm]
  \langle\phi|\phi\rangle
  &\ =\ \tr\bigl(E^{[1]}\cdots E^{[L]}\bigr)\ \ ,
\end{align}
  where
\begin{align}
  \bigl(E^{[k]}_X\bigr)_{\alpha\beta,\mu\nu}
  &\ =\ \sum_{s,\gamma,\rho}\bigl(\bar{A}^{[k]}\bigr)_{\alpha\mu}^s\bigl(A^{[k]}\bigr)_{\gamma\rho}^s\,\langle\gamma\rho|X|\beta\nu\rangle
\end{align}
  and $E^{[k]}=E^{[k]}_{\idop}$. A pictorial interpretation of the
  expectation value is provided in Figure \ref{fig:TransferMatrix}.
  The two traces can be evaluated by diagonalization of the transfer
  matrices $E^{[k]}$, considered as an operator mapping matrices on
  the auxiliary space $\cH_{(k,R)}$ to matrices acting on the
  auxiliary space $\cH_{(k,L)}$.\footnote{In other words: We read the
    operators $E$ from right to left.} In the thermodynamic limit and
  with $|i-j|\to\infty$, the only contribution will come from the
  highest eigenvalue. All other contributions are suppressed
  exponentially due to our mass gap assumption.
  It can easily be seen that the identity matrices on the auxiliary
  spaces are left and right ``eigenvectors'' with eigenvalues
  $\dim\cH_k/\dim\cH_{(k,L)}$ and $\dim\cH_k/\dim\cH_{(k,R)}$,
  respectively. Indeed, due to Schur's Lemma we have
\begin{align}
  E_{\alpha\beta,\mu\nu}^{[k]}\idop_{\mu\nu}
  \ &=\ C_R\,\idop_{\alpha\beta}
  \qquad\text{ and }\\[2mm]
  \idop_{\alpha\beta}\,E_{\alpha\beta,\mu\nu}^{[k]}
  \ &=\ C_L\,\idop_{\mu\nu}\ \ .
\end{align}
  Moreover it is obvious that
\begin{align}\nonumber
  \dim(\cH_k)
  \ &=\  \idop_{\alpha\beta}\,E_{\alpha\beta,\mu\nu}^{[k]}\,\idop_{\mu\nu}
  \ =\ C_R\,\tr_{\cH_{(k,R)}}(\idop)\\[2mm]
  \ &=\ C_R\,\dim(\cH_{(k,R)})\ \ ,
\end{align}
  and similarly for $C_L$. 
  \begin{comment}
  More formally, the previous equations
  should be written as
\begin{align}
  E^{[k]}|\idop_{\cH_{(k,R)}}\rangle
  \ =\ \frac{\dim\cH_k}{\dim\cH_{(k,R)}}\,|\idop_{\cH_{(k,L)}}\rangle
  \qquad\text{ and }\qquad
  \langle\idop_{\cH_{(k,L)}}|E^{[k]}
  \ =\ \frac{\dim\cH_k}{\dim\cH_{(k,L)}}\,\langle\idop_{\cH_{(k,R)}}|\ \ .
\end{align}
\end{comment}
  Since $E^{[k]}$ is a completely positive map it is guaranteed that
  there is no greater eigenvalue (Ref.\ \onlinecite{Paulsen2002}, Prop 3.6). On the other hand, due to
  our mass gap assumption this eigenvalue is non-degenerate, even in
  absolute value. \cite{Perez-Garcia:2007:MPS:2011832.2011833}

  In the limit of large separation, $|i-j|\gg1$, we can rewrite the
  desired expectation value in a factorized form as
\begin{align}\nonumber
  \langle\sigma_{ij}^{ab}\rangle
  \ &=\ \frac{\langle\idop_{\cH_{(i,L)}}|E_{J_L^a}^{[i]}|\idop_{\cH_{(i,R)}}\rangle}{\dim\cH_i}\,\frac{\langle\idop_{\cH_{(j,L)}}|E_{J_R^b}^{[j]}|\idop_{\cH_{(j,R)}}\rangle}{\dim\cH_j}\\[2mm]\label{eq:Factorization}
  \ &=\ \langle J_{i,L}^a\rangle\langle J_{j,R}^b\rangle \ \ .
\end{align}
  We note that the result still depends on the representation spaces
  describing the start and the end point of the original string.

% ***********************************************************************
\subsubsection*{Step 2: Employing Weyl symmetry}

  In the second part of the derivation we focus on the $a$ dependence
  of the expectation value $J^a:=\langle J_{i,L}^a\rangle$ which we
  claim to be proportional to $\Omega^{at}$ with
  $t=[\cH_{(i,R)}]$. The same reasoning can be used to derive that
  $\langle J_{j,R}^b\rangle \propto\ \Omega^{b[\cH_{(j,L)}]} =
  \Omega^{-bt}$ from which the main result, Eq.\ \eqref{eq:SOLimit},
  follows. Here we used the chain of equalities
  $[\cH_{(j,L)}]=[\cH_{(i,R)}^\ast]=-[\cH_{(i,R)}]=-t$.

  Since the operators $J_{i,L}^a$ contain Cartan elements only, their
  expectation value can be calculated most easily in an orthonormal
  basis $|\alpha\beta\rangle$ of the auxiliary space
  $\cH_{(i,L)}\otimes\cH_{(i,R)}$ which respects the weight space
  decomposition. In such a basis the operator $J^a_{i,L}$ is
  represented by a diagonal matrix with components
  $J^a_{\alpha\beta}$. In order to keep the notation simple we shall
  use the abbreviation $\alpha\in\mu$ if $|\alpha\rangle$ is contained
  in the weight space with weight $\mu$ (of $\cH_{(i,L)}$ in this
  case). Moreover, we wish to recall that the matrices
  $\left(A^{[k]}\right)^s_{\alpha\beta}$ are $SU(N)$ invariant
  projections from auxiliary space to physical space which can be
  represented as the matrix element $\langle
  s|\alpha\beta\rangle$. From the definition of the expectation values
  $J^a$ in Eq.\ \eqref{eq:Factorization} we immediately conclude
\begin{equation}
\begin{split}
  {J}^a
  \ &=\ \frac{1}{\dim\cH_i} \sum_{s,\alpha,\beta}\bigl|\langle
 s|\alpha\beta\rangle\bigr|^2 {J}^a_{\alpha\beta}\\[2mm]
  \ &=\ \frac{1}{\dim\cH_i}\sum_{\mu,\nu}\sum_{\substack{\alpha\in\mu,\beta\in\nu\\s\in\mu+\nu}}\bigl|\langle s|\alpha\beta\rangle\bigr|^2
 {J}^a_{\alpha\beta}\ \ .
\end{split}
\end{equation}
  In the second equality, instead of summing directly over all basis
  vectors, we first sum over weight spaces followed by a sum over
  vectors spanning a certain weight space. We also used an obvious
  selection rule for the weights entering the Clebsch-Gordan
  coefficients $\langle s|\alpha\beta\rangle$. The values
  ${J}^a_{\alpha\beta}$ do not directly depend on $\alpha$ and
  $\beta$, but only on the weight space they belong to. We may thus
  define
\begin{equation}
 {J}^a_{\mu\nu}
  \ :=\ {J}_{\alpha\beta}^a
  \qquad\text{ with }\qquad
 \alpha\in\mu,~\beta\in\nu\ \ .
\end{equation}
  Furthermore, it is convenient to introduce the abbreviation
\begin{align}
  \label{eq:P}
  P(\mu,\nu)
  \ :=\ \sum_{\substack{\alpha\in\mu,
      \beta\in\nu\\s\in\mu+\nu}}\bigl|\langle s|\alpha\beta\rangle\bigr|^2
\end{align}
  such that above expression can be written as
\begin{equation}\label{eq24}
  {J}^a
  \ =\ \frac{1}{\dim\cH_i} \sum_{\mu,\nu} P(\mu,\nu)
  {J}^a_{\mu\nu}\ \ .
\end{equation}
  At this point we split the sum into orbits with respect to the Weyl
  group. To be more precise, we simplify Eq.\ \eqref{eq24} by
  restricting the summation to those weights $\mu$ and $\nu$ such that
  their sum is in the fundamental Weyl chamber, $\mu+\nu\in P^+$. All
  the other terms are obtained using the action of the Weyl
  group. Since the weights at the boundary of $P^+$ are invariant
  under a subgroup of the Weyl group this leads to an overcounting
  which is compensated by dividing through the order of the stabilizer
  subgroup $\cW_{\mu+\nu}\subset\cW$. This procedure yields
\begin{align}
  {J}^a
  &\ =\ \frac{1}{\dim\cH_i}\sum_{\substack{\mu,\nu\\\mu+\nu\in P^+}}
        \frac{1}{|\cW_{\mu+\nu}|}\sum_{S\in\cW}
        P(S\mu,S\nu){J}^a_{S(\mu),S(\nu)}\nonumber\\[2mm]
  &\ =\ \frac{1}{\dim\cH_i}\sum_{\substack{\mu,\nu\\\mu+\nu\in P^+}}
        \frac{P(\mu,\nu)}{|\cW_{\mu+\nu}|}\sum_{S\in\cW}
        {J}^a_{S(\mu),S(\nu)}\nonumber\\[2mm]\label{eq:aux2}
  &\ =\ \frac{1}{\dim\cH_i}\sum_{\substack{\mu,\nu\\\mu+\nu\in P^+}}
        \frac{P(\mu,\nu)}{|\cW_{\mu+\nu}|}
  K^a_{\mu,\nu}\ \ ,
\end{align}
  In the second equation the Weyl invariance of $P(\mu,\nu)$ is used:
  $P(\mu,\nu)=P(S\mu,S\nu)$ for all $S\in\cW$. This is proven
  in Appendix \ref{ap:WeylGroupInvariance}. The third equation defines
  $K^a_{\mu,\nu}$. Since $P(\mu,\nu)$ is independent of $a$, we are
  left to show that $K^a_{\mu,\nu}(a)\propto\Omega^{a[\nu]} =
  \Omega^{at}$. The identification of $[\nu]$ with $t$ follows since
  the label $[\nu]$ is the same for all weights $\nu$ appearing in the
  decomposition of the $su(N)$ representation $\cH_{(i,R)}$. This is a
  direct consequence of the fact that the ground state $|\phi\rangle$
  was assumed to be in a well-defined topological phase.

% ***********************************************************************
\subsubsection*{\label{sc:JEval}Step 3: Weyl group gymnastics}

  At this point, all ingredients are set to show that
  $K^a_{\mu,\nu}\propto\Omega^{a[\nu]}$, where $K^a_{\mu,\nu}$ is
  defined by Eq.\ \eqref{eq:aux2}:
\begin{equation}
  K^a_{\mu,\nu}
  \ =\ \sum_{S\in\cW}S(\mu+\nu)(H^a)\,\Omega^{-S(\nu)(H^\rho)}\ \ .
\end{equation}
  Writing the weights as $\nu=\sum_lc_l\epsilon_l$ and
  $\mu+\nu=\sum_kd_k\epsilon_k$ (with the ``gauge fixing''
  $\sum_lc_l=\sum_kd_k=0$), respectively, and using the Weyl group
  action specified in Eq.\ \eqref{eq:WeylGroupAction} allows us to
  rewrite this expression in the form
\begin{align}\nonumber
  K^a_{\mu,\nu}
  \ =&\ \sum_{k}d_k\sum_{\sigma(k)=a}\Omega^\wedge\biggl(-\sum_lc_{l}\epsilon_{\sigma(l)}(H^\rho)\biggr) \\[2mm]\nonumber
     &-  \sum_{k}d_k\sum_{\sigma(k)=a+1}\Omega^\wedge\biggl(-\sum_lc_{l}\epsilon_{\sigma(l)}(H^\rho)\biggr)\\[2mm]
  \ =&\ \sum_k d_k\bigl(Q^{(a)}_k-Q^{(a+1)}_k\bigr)\ \ .
\end{align}
  Let us now focus on the sum over the different permutations $\sigma$
  which has been abbreviated by $Q_k^{(a)}$ in the previous
  formula. Our goal is to remove the constraint involving the index
  $a$ in the summation over the permutations and to convert it into an
  explicit dependence of the whole expression. This can be achieved by
  the following simple resummation which makes use of the cyclic
  permutations $\sigma_a$ and of Eq.\ \eqref{eq:aux4}
\begin{align}\nonumber
  Q_k^{(a)}
  \ =\ & \sum_{\sigma(k)=a}\Omega^\wedge\biggl(-\sum_lc_{l}\epsilon_{\sigma(l)}(H^\rho)\biggr)\\[2mm]\nonumber
  \ =\ & \sum_{\sigma(k)=N}\Omega^\wedge\biggl(-\sum_lc_{\sigma^{-1}(l)}\epsilon_{\sigma_a(l)}(H^\rho)\biggr)\\[2mm]\nonumber
  \ =\ & \sum_{\sigma(k)=N}\Omega^\wedge\biggl(-\sum_lc_{l}\epsilon_{\sigma(l)}(H^\rho)+a\sum_lc_l\biggr.\\[2mm]\nonumber
     & \quad\quad\quad\quad\quad\quad \biggl.-\sum_{l=N-a+1}^N Nc_{\sigma^{-1}(l)}\biggr)\\[2mm]
  \ =\ & \Omega^{a[\nu]}\,Q_k^{(N)}\ \ .
\end{align}
  The last expression arises from the following simplifications on the
  third row.
  The first sum in the exponent of the last equation is independent of
  $a$. The second sum vanishes due to our choice of ``gauge fixing''
  $\sum_lc_l=0$. The third sum contains the information we are
  after. Formula \eqref{eq:WeightConversion} implies the relation
  $-Nc_k=\sum_a a\nu_a=[\nu]$ (modulo $N$) for any index $k$. Hence
  each term in this sum is equal to $[\nu]=t$. Moreover, there are
  exactly $a$ of these terms in this sum. This gives the desired
  dependence of $K^a_{\mu,\nu} \propto \Omega^{a[\nu]}$ on the index
  $a$. The equation
\begin{equation}
  J^a\ =\ \frac{\Omega^{at}(1-\Omega^{t})}{\dim\cH_i}\sum_{\substack{\mu,\nu\\\mu+\nu\in P^+}}\frac{P(\mu,\nu)}{|\cW_{\mu+\nu}|} \sum_k d_kQ^{(N)}_k
\end{equation}
  follows immediately. We have thus confirmed that the string order
  parameter is given by Eq.\ \eqref{eq:SOLimit} and that it is a
  suitable tool for measuring the topological phase of a state on a
  spin chain. Moreover, the previous equation also implies that the
  string order operator always has a vanishing expectation value as
  long as the edge modes transform according to a linear representation
  of $PSU(N)$, i.e.\ when the system is in a topologically trivial
  phase with $t=0$.

% ***********************************************************************
% ***********************************************************************
\subsection{\label{sc:proplim}Properties}

  The string order parameter that is derived from Eq.\
  \eqref{eq:StringOrderOp} has a number of desired features that one
  expects for a quantity capable of measuring a topological
  property. First of all, the factorization \eqref{eq:Factorization}
  implies the invariance under arbitrary block renormalization between
  the end points in questions. From a mathematical perspective this is
  the analogue of invariance under continuous deformations or choice
  of metric. Even though the factorized expression resembles a local
  correlation function one should bear in mind that the invariance of
  the ground state under $PSU(N)$ leads to a subtle entanglement which
  propagates from site to site and cannot be removed by block
  renormalization. \cite{Chen:PhysRevB.83.035107}

  It should be emphasized that the integer number $t$ associated with
  our string order parameter \eqref{eq:SOLimit} gives a reliable
  answer about the precise type of the topological phase. In contrast,
  entanglement entropies and spectra only encode information about the
  number of massless edge modes but not (at least not directly) about
  their representation type (see e.g.\ Ref.\ 
  \onlinecite{Tu:PhysRevB.80.014401,Pollmann:2012arXiv1204.0704P}). Indeed,
  even when only considering {\em irreducible} representations of
  $SU(N)$, the dimension is not sufficient to distinguish between
  a representation and its dual for instance. A systematic search for
  even more convincing examples already succeeds for $SU(3)$: This
  group has four different 15-dimensional irreducible representations
  labeled by $(2,1)$ and $(4,0)$ as well as their conjugates. While
  $(2,1)$ and $(4,0)$ belong to the class $[1]\in\Integer_3$,
  the representations $(1,2)$ and $(0,4)$ belong to the class
  $[2]\in\Integer_3$. So, even when forgetting about the possibility
  to form direct sums of irreducible representations we recognize that
  the dimension of a representation alone might not be sufficient to
  specify the topological phase it is associated with.

  The formula we derived for the string order and its interpretation
  in a sense assumes an ideal measurement. The form of the outcome and
  the particular dependence of the complex phase factor on the label
  $a$ rely on a very specific and fixed choice of basis for the Cartan
  generators. In a real physical measurement in a laboratory one will
  generally measure the expectation value for a linear combination of
  operators which slightly deviates from $H^a$. A more detailed
  analysis of this effect, just as of finite size corrections, is
  beyond the scope of the present article.

% ***********************************************************************
% ***********************************************************************
% ***********************************************************************
\section{\label{sc:Applications}Numerical verification}

  In this Section, it will be verified in a concrete physical setup
  that the string order parameter defined in Section
  \ref{sc:StringOrder} is capable of measuring the topological order
  of a spin chain. For this purpose we define a family of $PSU(3)$
  invariant Hamiltonians which smoothly interpolates between two
  distinct topologically non-trivial phases. We determine the ground
  states numerically using DMRG and study the behavior of the string
  order parameter and its associated topological order
  parameter~$t$. The numerical results clearly confirm our theoretical
  predictions. The {\em complex phase} of the string order parameter
  is quantized and jumps at the phase transition.

% ***********************************************************************
% ***********************************************************************
\subsection{Setup and idea}

  In what follows, we shall consider a family of $PSU(3)$ invariant
  spin chains with periodic boundary conditions. The on-site Hilbert
  spaces are all chosen to be equal to the eight-dimensional adjoint
  representation of $SU(3)$, which is described by the highest weight
  $(1,1)$. Since Eq.\ \eqref{eq:SUClass} implies $[(1,1)]\equiv0$,
  this is clearly a representation of $PSU(3)$. We start with a
  discussion of two particular states $|\phi_1\rangle$ and
  $|\phi_2\rangle$ and their associated parent Hamiltonians $H_1$ and
  $H_2$. For these two systems we have full analytical control over
  all relevant properties such as the energy gap and the topological
  phase. We then consider the family of Hamiltonians
\begin{align}
  \label{eq:Family}
  H(c)\ =\ cH_1+(1-c)H_2
  \quad\text{ with }\quad
  c\in[0,1]\ \ .
\end{align}
  Our basic idea is to determine the ground state and the string order
  parameter numerically as a function of $c$. Since, however, the
  structure of the Hamiltonian $H(c)$ is quite complicated we will
  instead implement the numerics using a truncated version
  $H_\text{trunc}(c)$ which exhibits the same qualitative behavior.

  The state $|\phi_1\rangle$ is a matrix product state defined as
  follows: As the left and right auxiliary spaces we choose the two
  distinct three-dimensional representations $\bar{3}$ and $3$ of
  $SU(3)$, with highest weight $(0,1)$ and $(1,0)$, respectively. The
  matrices $A$ correspond to the $SU(3)$ invariant projections
  $A:\bar{3}\otimes3\rightarrow8$ as described in Section
  \ref{sc:Preliminaries}. By construction, the state $|\phi_1\rangle$
  resides in the non-trivial topological phase $t=[(1,0)]=1$. As is
  well known, the parent Hamiltonian for an open chain of this form
  will lead to massless boundary spins transforming in the
  representations $\bar{3}$ and $3$, respectively. With periodic
  boundary conditions however, we end up with a unique ground state. A
  state which belongs to the topological class $t=1$ necessarily
  breaks inversion symmetry since the representations $3$ and
  $\bar{3}$ in the auxiliary space need to be treated on a different
  footing. Since the ground state is required to be non-degenerate,
  this actually provides an interesting challenge for the construction
  of a suitable two-site Hamiltonian as will be discussed below.

  The state $|\phi_2\rangle$ is obtained from $|\phi_1\rangle$ by
  inversion. In particular, the left auxiliary space of each site is
  interchanged with the right auxiliary space. As should be clear from the
  exchange of auxiliary spaces, the new state $|\phi_2\rangle$ resides
  in the non-trivial topological phase $t=[(0,1)]=2$. Of course we can
  also apply the inversion to the Hamiltonian $H_1$, resulting in a
  new Hamiltonian $H_2$ of which $|\phi_2\rangle$ is the unique ground
  state.

% ***********************************************************************
% ***********************************************************************
\subsection{\label{sc:Family}A family of Hamiltonians}

  We are now making the preceding statements more
  explicit, following the standard strategy of the AKLT construction.
  \cite{Affleck:PhysRevLett.59.799,Affleck:1987cy} Our goal
  is to find concrete expressions for the Hamiltonians $H_1$ and $H_2$
  as well as for the interpolating Hamiltonian $H(c)$ defined in
  \eqref{eq:Family}. This requires introducing the concept of Casimir
  operators (see also Appendix \ref{ap:CasimirOperators}) and the
  calculation of a few tensor products. It turns out that we can
  restrict our attention to Hamiltonians involving nearest neighbor
  interactions only.

\begin{figure}[t]
\includegraphics[width=\columnwidth]{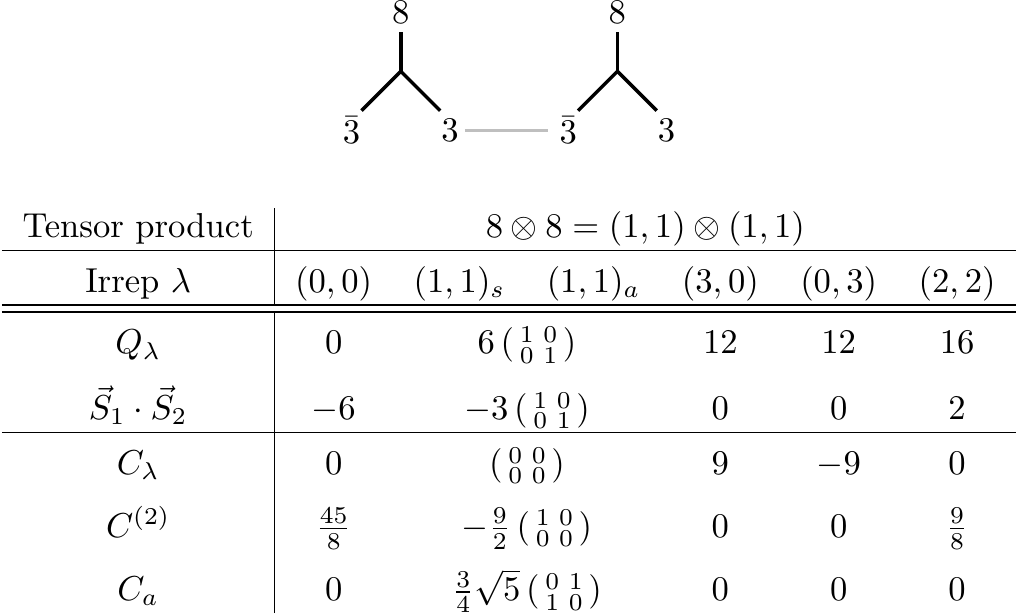}
  \caption{\label{fig:HBulk}Construction of the two-site
    Hamiltonian.}
\end{figure}

  The two-site Hilbert space decomposes as follows,
\begin{equation}
\begin{split}
  \label{eq:TPBulk}
  (1,1)\otimes(1,1)
  &\ =\ (0,0)\oplus(1,1)_s\oplus(1,1)_a\\[2mm]
  &\qquad\quad\oplus(3,0)\oplus(0,3)\oplus(2,2)\ \ .
\end{split}
\end{equation}
  The subscripts in $(1,1)_s$ and $(1,1)_a$ refer to the symmetric
  and to the anti-symmetric part of the tensor product. Schur's Lemma
  implies that $su(3)$ invariant Hamiltonians cannot change the type
  of representation. This leaves one parameter for each of the
  representations which occur with multiplicity one but four
  parameters for the representation $(1,1)$ which appears with
  multiplicity two. The latter can be thought of as the entries of a
  $2\times2$ matrix which acts on the multiplicity space of the
  representation $(1,1)$. In total, there is thus an eight-dimensional
  space of two-body Hamiltonians which commute with the action of
  $su(3)$. In what follows, we will express these explicitly in terms
  of invariant combinations of the spin operators $\vec{S}_1$ and
  $\vec{S}_2$ on the two sites.

  The basic objects we have at our disposal are the
  expression $Q_{12}=\vec{S}_1\cdot\vec{S}_2$ which is related to the
  quadratic Casimir $(\vec{S}_1+\vec{S}_2)^2$ as well as the cubic
  terms $C_{112}=d_{rst}S_1^rS_1^sS_2^t$ and
  $C_{122}=d_{rst}S_1^rS_2^sS_2^t$ which are defined using a symmetric
  invariant rank three tensor $d_{rst}$, see Appendix
  \eqref{ap:CasimirOperators}. In addition, we need to consider
  polynomials in these objects, potentially with permutations in the
  order of the operators. One example for such an operator would be
\begin{align}
  C^{(2)}
  \ :=\ d_{rst}d_{uvw}S_1^rS_1^uS_1^vS_2^wS_2^sS_2^t\ \ .
\end{align}
  A careful analysis shows that the eight-dimensional space of
  invariant operators acting on the tensor product $(1,1)\otimes(1,1)$
  is spanned by
\begin{equation}
\begin{split}
  &\bigl\langle
  \,1\,,\,
  Q_{12}\,,\,
  Q_{12}^2\,,\,
  Q_{12}^3\,,\,
  C_s=C_{112}+C_{122}\,,\,\\[2mm]
  &\quad C_a=C_{112}-C_{122}\,,\,
  C^{(2)}\,,\,
  [C_a,C^{(2)}]\,
  \bigr\rangle\ \ .
\end{split}
\end{equation}
  The action of some of these operators on the constituents of the
  tensor product \eqref{eq:TPBulk} is summarized in Figure
  \ref{fig:HBulk}. Note that $C_a$ is an operator which exchanges the
  symmetric and the anti-symmetric part of the tensor product. After
  some linear algebra, it turns out that a good choice for the
  interpolating two-site Hamiltonian entering \eqref{eq:Family} is
  given by\footnote{When making this specific choice we used input
    about the desired boundary modes (see below) in order to fix the
    projection in the two-dimensional multiplicity space of $(1,1)$.}
\begin{equation}
\begin{split}
  H(c)
  &\ =\ 1+\frac{9}{56}\vec{S}_1\cdot\vec{S}_2
       -\frac{5}{112}(\vec{S}_1\cdot\vec{S}_2)^2 \\[2mm]\label{eq:HBulk}
  &\hspace{0mm}-\frac{1}{112}(\vec{S}_1\cdot\vec{S}_2)^3
       +(1-2c)\frac{2}{7}C_a
       -\frac{4}{63}C^{(2)}\ \ .
\end{split}
\end{equation}
  We note that the deformation parameter $c$ only multiplies the term
  $C_a$ which explicitly breaks inversion symmetry. It is not obvious
  at all, but an explicit calculation shows that the Hamiltonian above
  reduces to a projector for $c=0$ and for $c=1$ (see the table in
  Figure \ref{fig:HBulk}). In both cases it projects onto the subspace
  generated by $(3,0)\oplus(0,3)\oplus(2,2)$ as well as two
  (different) one-dimensional subspaces in the
  two-dimensional multiplicity space of $(1,1)_s\oplus(1,1)_a$. The
  latter single out a specific copy of $(1,1)$ inside of
  $(1,1)_s\oplus(1,1)_a$. In other words, the space of zero-energy
  states (for two sites) is given by $(0,0)$ and states in a
  {\em complementary} copy of $(1,1)$ within
  $(1,1)_s\oplus(1,1)_a$ for $c=0$ and $c=1$. This is precisely the
  content of $(1,0)\otimes(0,1)$, i.e.\ the contribution of the four
  auxiliary sites with the singlet constraint imposed, thus showing
  that the Hamiltonians $H(0)$ and $H(1)$ are of AKLT-type.

  Since the numerical evaluation of the Hamiltonian \eqref{eq:HBulk}
  is quite time-consuming we shall henceforth work with the following
  family of truncated Hamiltonians,
\begin{align}
  \label{eq:FamilyTrunc}  
  H_{\text{trunc}}(c)
  &\ =\ 1+\frac{9}{56}\vec{S}_1\cdot\vec{S}_2
        +(1-2c)\frac{2}{7}C_a\ \ .
\end{align}
  In view of the structural similarity with the Hamiltonian
  \eqref{eq:HBulk} we believe that both share the same qualitative
  features. Evidence for this assertion comes from the exact
  diagonalization on a chain of $L=6$ sites.

% ***********************************************************************
% ***********************************************************************
\subsection{\label{sc:Numerics}Evaluation of the topological order
  parameter and discussion}

\begin{figure}[t]
\includegraphics[width=\columnwidth]{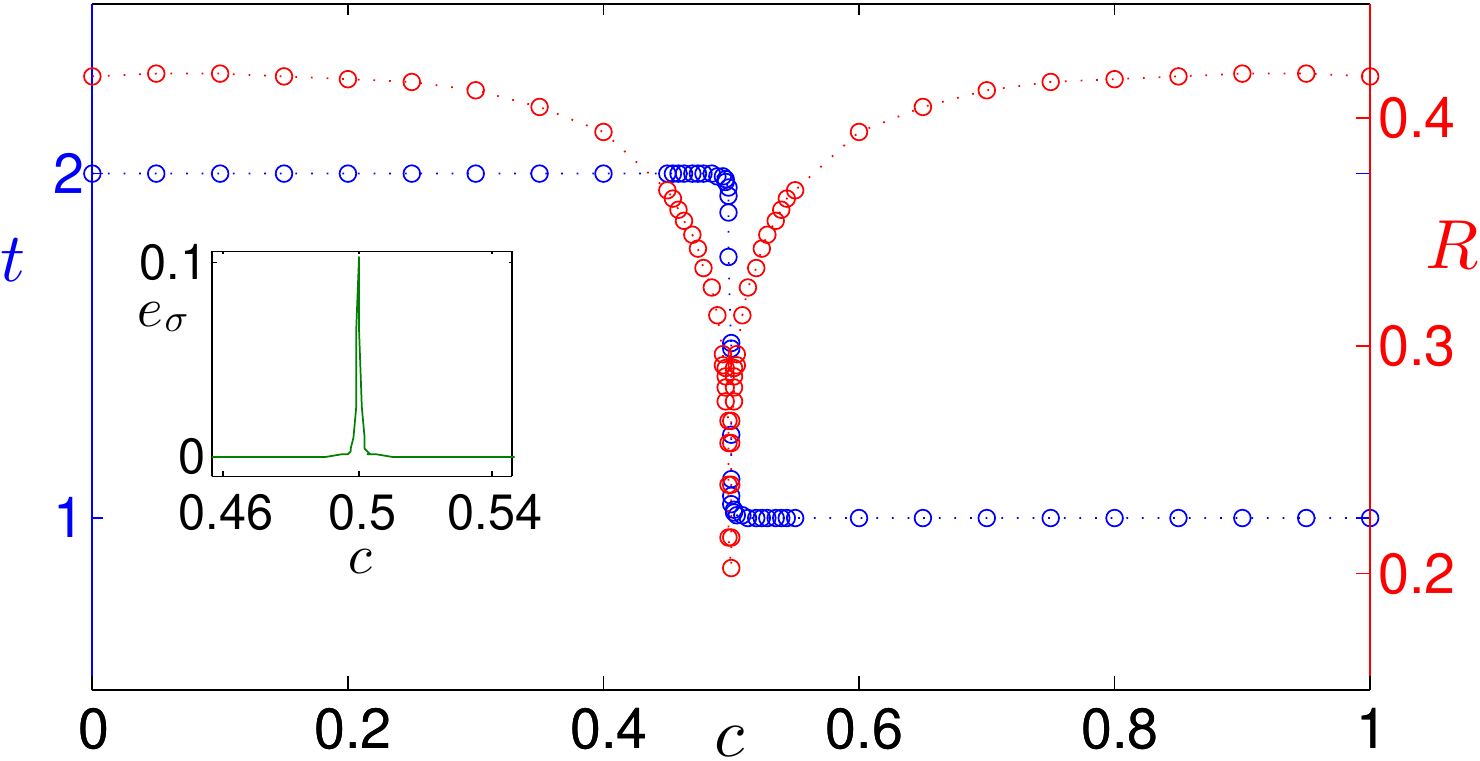}
  \caption{\label{fig:res}(Color online) A phase
    transition between two topological phases. The
    diagram shows the $c$-dependence of the parameters
    $t$ (blue, squares) and $R$ (red, circles). The inlet shows the
    deviation of the measured form of the string order matrix from its
    analytical form. We used adapted increments in the vicinity of the
    phase transition.}
\end{figure}

  For different values of $c$ in Eq.\ \eqref{eq:FamilyTrunc}, we have
  calculated the ground state using DMRG
  techniques.\cite{schollwoeck-2011-326} We have considered a chain of
  length $L=20$ and worked with an auxiliary space of dimension
  $D=400$. We calculated the expectation value of the
  string order parameter $\langle\sigma_{ij}^{ab}\rangle$ numerically,
  for the specific sites $i=5$ and $j=15$. We compared the resulting
  matrix to the expression
\begin{align}
  \label{eq:SU3Result}
  \overline{\langle\sigma_{ij}\rangle}
  \ =\ -R\mat1&\Omega^{2t}\\\Omega^t&1\tam
  \quad\text{ with }\quad
  \Omega = \exp\frac{2\pi i}{3}\ \ ,
\end{align}
  which is the theoretical prediction for the string order parameter
  in the limit of an infinite chain (see Eq.\ \eqref{eq:SOLimit}). The
  numerical values of the parameters $R$ and $t$ have been estimated
  by minimizing $e_\sigma=\tr(d\sigma\cdot d\sigma^\dagger)$, with $d\sigma =
  \langle\sigma_{ij}^{ab}\rangle -
  \overline{\langle\sigma_{ij}^{ab}\rangle}$. The results are plotted
  in Figure \ref{fig:res} and they are in perfect agreement with the
  theory. The parameter $t$ is quantized and restricted to the numbers
  $1$ and $2$, thus providing the desired label for the topological
  class of the system. Moreover, this parameter changes
  discontinuously at the value $c=1/2$.

  The failure of finding $R=0$ at the phase transition is probably due
  to finite bond dimension and finite system size. Indeed, apart from
  potential numerical deficiencies there are finite size corrections
  which have been neglected in the derivation of Eq.\
  \eqref{eq:SOLimit}. These finite size effects become more important
  as the mass gap goes to zero and the correlation length
  increases. Let us summarize two observations which provide evidence
  for this assertion. First of all, the error bars in Figure
  \ref{fig:res} which quantify the discrepancy of the numerical result
  from the analytical expression \eqref{eq:SU3Result} grow
  significantly close to the transition point. In addition, we
  compared the numerical results for $R$ at $c=1/2$ using two
  different bond dimensions $D=200$ and $D=400$. The drop from
  $R=0.31$ to $R=0.20$ is another signal of finite size effects.

  Of course, the transfer matrix method allows to compute
  the string order exactly, even for finite size of the system, once
  the eigenvalues and the eigenvectors of the transfer matrix have
  been determined. However, our numerical analysis here should merely
  be regarded as a proof of principle. A more accurate treatment will
  be left for future work. Despite our numerical limitations we still
  clearly see the crossover from one topological phase to another.

  In addition to the previous investigations we applied the same
  method to the full parent Hamiltonians $H_1$ and $H_2$. Also in
  this case, the numerical analysis confirmed our analytical
  expectation that the corresponding ground states belong to the
  non-trivial topological classes $1$ and $2$, respectively.

% ***********************************************************************
% ***********************************************************************
% ***********************************************************************
\section{Conclusions}

  In our paper, we have searched for a physical observable which
  allows to distinguish the $N$ different topological phases of
  $PSU(N)$ spin chains. To achieve this goal we have proposed a
  non-local string order operator in equation \eqref{eq:StringOrderOp}
  and we have shown that its expectation value provides an unambiguous
  measure for the topological phase the chain resides in. In essence,
  our string order parameter extracts the projective class of the
  representations according to which potential (virtual) massless
  boundary modes transform in. It should be emphasized that, in
  contrast to earlier studies, our string order parameter is matrix
  valued. All matrix entries are equal in absolute value and identical
  to zero in the topologically trivial phase. The
  information about the -- quantized -- topological phase of the chain
  is contained in the relative complex phases between different
  matrix entries. More precisely, the quotient of two suitably chosen
  matrix elements is completely sufficient in order to extract the
  quantized topological order parameter determining the topological
  phase.  Our analytical results are supported by the numerical study
  of a family of $PSU(3)$ Hamiltonians which interpolates between two
  distinct non-trivial topological phases. Since the realization of
  these two phases enforces the breaking of inversion symmetry, the
  Hamiltonian employs a new construction scheme making explicit use of
  higher order Casimir operators. We find full agreement between our
  analytical predictions and the numerical results. Indeed, Figure
  \ref{fig:res} clearly exhibits a robust quantization of the
  topological order parameter.
  
  Even though tentative results have been included here, we believe
  that $SU(N)$ spin chains deserve further numerical study. First of
  all, our numerical investigation of the string order parameter only
  covered a special family of $SU(3)$ spin chains, the interpolation
  between two topologically non-trivial phases. While this provided
  the desired proof of principle that our method works in practice,
  one could similarly analyze the behavior of the string order
  parameter when interpolating between a non-trivial phase and the
  trivial phase. An important open problem in this context is the
  identification of the type of phase transitions that occur when
  crossing the boundary between two distinct topological phases. For
  our model Hamiltonian \eqref{eq:FamilyTrunc} we analyzed the gap behavior
  in the vicinity of the transition point $c=1/2$. However, at this
  point of time our DMRG results are not accurate enough to be able to
  draw a final conclusion. Another possible avenue to uncover the
  nature of the phase transition is the investigation of the scaling
  behavior of the entanglement
  entropy.\cite{Korepin:PhysRevLett.92.096402,Calabrese:2004JSMTE..06..002C}
  The latter is directly accessible from the DMRG representation of
  the ground state. However, just as before accurate results would
  require increasing bond dimension and system
  size.\cite{Fuhringer2008:AnnPhys.17.922}

  Another natural direction is the extension of our numerical study
  to larger values of $N$. Since $PSU(N)$ spin chains have $N$
  distinct topological phases, we expect a complicated phase diagram
  with a large number of different phase transitions which might be
  implemented. It would be interesting to investigate whether each
  pair of mutually distinct phases is directly connected or whether
  they are only connected via a series of phase transitions each of
  which changes the $\Integer_N$ topological order by one unit for
  instance.

  It is evident that systems which are invariant under continuous
  symmetries different than $SU(N)$ should also admit a string order
  parameter similar to the one described in the current paper. Even
  though the groups based on $SU(N)$ are the most interesting ones due
  to the large size of their center, it is known
  \cite{Duivenvoorden:2012arXiv1206.2462D} that two and three
  distinct {\em non-trivial} topological phases, respectively, also
  exist for the symmetry groups $E_6$ and $Spin(2N)$ (the universal
  cover of $SO(2N)$). Just as for $PSU(N)$ a single expectation value
  will not be sufficient to distinguish between different types of
  topological order for such symmetries. In addition, an extension to
  certain classes of supersymmetric or anisotropic systems looks
  feasible. It should be noted, however, that the respective
  symmetries of these systems are described by supergroups or quantum
  groups and that a classification of topological phases is still
  missing in that context. Nevertheless, it seems likely that our
  formula \eqref{eq:StringOrderOp} will be applicable in anisotropic
  spin chains with $SU_q(N)$ quantum group symmetry without
  modification.

  It remains to be clarified how our string order parameter relates to
  other recent proposals for the determination of the projective class
  of (virtual) edge
  modes.\cite{Haegeman:1201.4174v1,Pollmann:2012arXiv1204.0704P} While
  there is no fundamental obstruction in applying these techniques to
  the case of $PSU(N)$, the details still need to be worked out. In
  particular, we would like to remark that both Refs.\
  \onlinecite{Haegeman:1201.4174v1,Pollmann:2012arXiv1204.0704P}
  adopt a perspective which is somewhat different from ours: Their
  discussion is based on relations between discrete group
  elements (possibly interpreted as elements of subgroups of a
  continuous group), while our proposal only features the underlying
  Lie algebra and, in fact, only its abelian part. As a result, our
  final formula \eqref{eq:StringOrderOp} for the string order
  parameter is easy to evaluate on the standard basis of the spin
  states. This statement is independent of whether the ground state is
  represented as a matrix product state or not.

  Let us finally address an interesting conceptual issue that arises
  in connection with our work. For the original $SU(2)$ AKLT chain it
  is well known that the existence of a non-trivial Rommelse-Den\,Nijs
  string order \cite{DenNijs:PhysRevB.40.4709} is equivalent to the
  breaking of a discrete hidden symmetry $\Integer_2\times\Integer_2$.
  \cite{Kennedy:PhysRevB.45.304,Kennedy:10.1007/BF02097239,Oshikawa:0953-8984-4-36-019,Totsuka:0953-8984-7-8-012}
  This intimate relationship can be made manifest by means of a {\em
    non-local} transformation of the spin chain. It would be very
  interesting to investigate whether a similar relationship exists for
  general $SU(N)$ spin chains and to analyze the symmetry breaking
  patterns of discrete groups that arise in this way when considering
  the full hierarchy of topological phases.
  \cite{Duivenvoorden:2012arXiv1206.2462D} The relationship between
  string order and discrete hidden symmetries for higher rank groups
  was also discussed in Ref.~\onlinecite{Tu:2008JPhA...41O5201T}.

% ***********************************************************************
% ***********************************************************************
% ***********************************************************************
\subsubsection*{Acknowledgment}

  We gratefully acknowledge useful discussions with A.\ Alex, J.\ von
  Delft, M.\ Kalus, K.\ Rodriguez, A.\ Rosch, Z.\ Shaikh, S.\ Trebst
  and M.\ Zirnbauer. The work of Kasper Duivenvoorden is funded by the
  German Research Foundation (DFG) through the SFB$|$TR\,12
  ``Symmetries and Universality in Mesoscopic Systems'' and the
  ``Bonn-Cologne Graduate School of Physics and Astronomy''
  (BCGS). The work of Thomas Quella is funded by the DFG through
  Martin Zirnbauer's Leibniz Prize, DFG grant no.\ ZI 513/2-1.

\appendix
% ***********************************************************************
% ***********************************************************************
\section{\label{ap:WeylGroupInvariance}Weyl group invariance of
  Clebsch-Gordan coefficients}

  In this Appendix it will be shown that the expression $P(\mu,\nu)$
  defined in Eq.\ \eqref{eq:P}, is invariant under a Weyl
  transformation of the weights $\mu$ and $\nu$.  Note that the
  expression can be rewritten as a trace over three orthogonal
  projections:
\begin{equation}
  P(\mu,\nu)
  \ =\ \sum_{\substack{i\in\mu,j\in\nu\\s\in\mu+\nu}}\bigl|\langle s|ij\rangle\bigr|^2
  \ =\ \tr(\Pi_{\mu\nu} \Pi_\cH \Pi_{\mu\nu})\ \ .
\end{equation}
  Recall that $\cH_k\subset \cH_{(k,L)}\otimes \cH_{(k,R)}$: $\Pi_\cH$
  denotes the orthogonal projection onto this subspace. $\Pi_{\mu\nu}$
  denotes the orthogonal projections on the weight space
  $V_\mu\otimes V_\nu\subset \cH_{(k,L)}\otimes \cH_{(k,R)}$.

  The Weyl group is not only the symmetry group of the root system,
  but it can also be defined as the quotient group of the normalizer
  of the maximal torus with the centralizer of the maximal torus:
  $\cW=N(T)/Z(T)$. The maximal torus of $SU(N)$ simply consists of all
  diagonal matrices with elements of $U(1)$ on the diagonal and
  determinant 1. Just like elements in the Cartan subalgebra $\h^*$,
  elements of the maximal torus have a simple action on states
  $v_\lambda$ with a well defined weight $\lambda$. For $h = \exp{H}$
  ($H\in\h$) one simply obtains
\begin{equation}
  \rho(h)v_\lambda
  \ =\ \exp{\lambda(H)} v_\lambda\ \ .
\end{equation}
  Let $\sigma: SU(N)\supset N(T)\rightarrow
  \cW\rightarrow \text{Aut}(\Gamma_V)$, where $\Gamma_V$ is the space
  of weights appearing in the representation $V$. Explicitly,
  $\sigma_w\mu(h) = \mu(w^{-1}hw)$. Weyl invariance of $P(\mu,\nu)$
  will follow from
\begin{equation}
  P(\sigma_w\mu,\sigma_w\nu)
  \ =\ P(\mu,\nu)\ \ .
\end{equation}
  The advantage of this approach is that since $w\in SU(N)$, the
  action of the Weyl group is trivial to implement on $V_1\otimes
  V_2$. Denote $\rho_i:SU(N)\rightarrow V_i$ for $i\in\{1,2\}$. Using
  this action we aim to show that:
\begin{equation}\label{eq:aux1}
  \rho_i(w)\Pi_\mu\rho_i(w)^{-1}
  \ =\ \Pi_{\sigma_w\mu}\ \ .
\end{equation}
  Since if this holds, the Weyl transformed function $P$ can be
  rewritten as
\begin{align}\nonumber
  &P(\sigma_w\mu,\sigma_w\nu)  = \\[2mm]\nonumber
  &\ =\ \tr( \Pi_{\sigma_w\mu\sigma_w\nu}  \Pi_\cH \Pi_{\sigma_w\mu\sigma_w\nu})\\[2mm]\nonumber
  &\ =\ \tr( \rho_{12}(w)\Pi_{\mu\nu}\rho_{12}(w)^{-1} \Pi_\cH\rho_{12}(w)\Pi_{\mu\nu}\rho_{12}(w)^{-1})\\[2mm]\nonumber
  &\ =\ \tr( \Pi_{\mu\nu} \rho_{12}(w)^{-1} \Pi_\cH \rho_{12}(w)  \Pi_{\mu\nu}) \\[2mm]
  &\ =\ \tr( \Pi_{\mu\nu}  \Pi_\cH \Pi_{\mu\nu})
   \ =\ P(\mu,\nu)\ \ , 
\end{align}
  which shows that $P(\mu,\nu)$ is Weyl invariant. In the second equality
  $\rho_{12} = \rho_1\otimes\rho_2$. In the third equality we make use
  of the cyclic property of the trace to cancel the outer two maps
  $\rho_{12}(w)$ and $\rho_{12}(w)^{-1}$. In the fourth equality we
  make use of the fact that $\rho_{12}(w)$ and $\Pi_\cH$ commute. We
  are left to check the validity of Eq.\ \eqref{eq:aux1}. Let $v_\mu\in
  V_\mu$ and let $h$ be an element in the maximal torus. The chain of
  equalities
\begin{equation}
\begin{split}
  \rho(h)\rho(w)v_\mu
  &\ =\ \rho(w) \rho(w^{-1}hw)v_\mu\\[2mm]
  &\ =\ {\mu(w^{-1}hw)}\rho(w)v_\mu\\[2mm]
  &\ =\ {\sigma_w\mu(h)}\rho(w)v_\mu
\end{split}
\end{equation}
  shows that $\rho(w)v_\mu\in V_{\sigma_w\mu}$. From this,
  Eq.\ \eqref{eq:aux1} follows.

% ***********************************************************************
% ***********************************************************************
\section{\label{ap:CasimirOperators}Casimir operators of
  $\mathbf{su(3)}$}

  The Casimir elements of a Lie algebra are polynomials in its
  generators $S^r$ which are central, i.e.\ which commute with each of
  the generators. For $su(3)$ there are two algebraically independent
  Casimir operators. One is the usual square of the spin vector
  $\vec{S}^2$. It is associated with a non-degenerate invariant form
  and can be expressed as $\vec{S}^2=\kappa_{rs}S^rS^s$ where
  $\kappa_{rs}$ is an invariant symmetric rank two tensor. The second
  Casimir is a cubic invariant
  $(\vec{S},\vec{S},\vec{S})=d_{rst}S^rS^sS^t$ which can be
  constructed from a non-vanishing invariant symmetric rank three
  tensor $d_{rst}$. Up to normalization, all invariant tensors of
  $su(3)$ are obtained by choosing suitable representations and by
  considering traces of the form 
\begin{align}
  \label{eq:Tensors}
  t^{a_1\cdots a_n}
  \ =\ \tr(S^{a_1}\cdots S^{a_n})\ \ .
\end{align}
  These tensors are not all independent. On the contrary, there exist
  algebraic relations between the tensors which may be used to reduce
  higher rank tensors to those of relatively low degree.

  For $su(3)$ the most convenient way of finding explicit expressions
  for the tensors \eqref{eq:Tensors} is to employ the fundamental
  representation in which the spin operators $S^r=\lambda^r/2$ are
  proportional to the Gell-Mann matrices $\lambda^r$ (see e.g.\ Ref.\ 
  \onlinecite{Greiter:PhysRevB.75.184441}). One then defines
\begin{align}
  \kappa^{rs}
  \ &=\ \tr(\lambda^r\lambda^s)
   \ =\ 2\delta^{rs}\ , \\[2mm]
  d^{rst}
  \ &=\ \frac{1}{4}\tr\bigl(\{\lambda^r,\lambda^s\}\lambda^t\bigr)\ \ .
\end{align}
  By construction, $\kappa^{rs}$ and $d^{rst}$ are manifestly
  symmetric. The matrices $\kappa^{rs}$ and its inverse,
  $\kappa_{rs}=\delta_{rs}/2$, serve as a metric which can be used to
  raise and lower indices, just as in special and in general
  relativity. The tensors which are used for the construction of the
  Casimir operators are $\kappa_{rs}$ and
  $d_{rst}=\kappa_{ru}\kappa_{sv}\kappa_{tw}d^{uvw}$.

  Since Casimir operators commute with the action of $su(3)$, they
  are represented as scalars on irreducible representations. With our
  normalization conventions, the eigenvalues of the quadratic and the
  cubic Casimir operator,
\begin{align}
  \label{eq:CasimirEigenvalues}
  Q\ =\ 4\,\kappa_{rs}S^rS^s
  \quad\text{ and }\quad
  C\ =\ 8\,d_{rst}S^rS^sS^t\ \ ,
\end{align}
  on the irreducible representation with highest weight $\lambda$ are
  given by
\begin{align}\nonumber
  Q_\lambda
  &=\ (\lambda,\lambda+2\rho) \\[2mm]
   &=\ \frac{2}{3}\bigl(\lambda_1^2+\lambda_2^2+\lambda_1\lambda_2+3\lambda_1+3\lambda_2\bigr)
  \quad\text{ and }\\[2mm]\nonumber
  C_\lambda
  &=\
  \frac{1}{2}(\lambda_1-\lambda_2)\biggl[\frac{2}{9}(\lambda_1+\lambda_2)^2+\frac{1}{9}\lambda_1\lambda_2+\lambda_1+\lambda_2+1\biggr].
\end{align}
  In contrast to $Q_\lambda$, the cubic Casimir $C_\lambda$ can
  distinguish between a representation $\lambda=(\lambda_1,\lambda_2)$
  and its dual $\lambda^+=(\lambda_2,\lambda_1)$. We also see that
  $C_\lambda$ vanishes on all representations which are self-dual.

%Merlin.mbs v4.21 2009-07-09.
%\bibliography{../bibliographyKasper}
%\bibliography{bibarticle2b}

%Merlin.mbs v4.21 2009-07-09.
%

\end{document}